\documentclass[amsmath,amssymb,aps,pra]{revtex4}
\usepackage{bm}
\usepackage[dvips]{graphicx}
\usepackage{epsfig,bm}

\begin{document}

\title{Cooperative scattering and radiation pressure force in dense atomic clouds}

\author{R. Bachelard$^{a}$, N. Piovella$^{b}$, and Ph. W. Courteille$^{c}$}

\affiliation{$^{a}$ University of Nova Gorica, School of Applied
Sciences, Vipavska 11c SI-5270 Ajdovšcina, Slovenia\\
    $^{b}$Dipartimento di Fisica, Universit\`a Degli Studi di Milano, Via Celoria 16, I-20133 Milano, Italy\\
    $^{c}$Instituto de F\'{i}sica de S\~{a}o Carlos, Universidade de S\~{a}o Paulo, 13560-970 S\~{a}o Carlos, SP, Brazil.}

\date{\today}

\begin{abstract}
Atomic clouds prepared in 'timed Dicke' states, i.e. states where
the phase of the oscillating atomic dipole moments linearly varies
along one direction of space, are efficient sources of
superradiant light emission~[M.O. Scully et al., Phys. Rev. Lett.
\textbf{96}, 010501 (2006)]. Here, we show that, in contrast to
previous assertions, timed Dicke states are \emph{not} the states
automatically generated by incident laser light. In reality, the
atoms act back on the driving field because of the finite
refraction of the cloud. This leads to non-uniform phase shifts
which, at higher optical densities, dramatically alter the
cooperative scattering properties, as we show by explicit
calculation of macroscopic observables, such as the radiation
pressure force.
\end{abstract}
\maketitle

\section{Introduction}

The fact that ensembles of point-like scatterers respond
collectively to incident radiation is well-known since the seminal
paper of Dicke \cite{Dicke}. The collective phenomenon, termed
superradiance, has been the topic of a huge amount of theoretical
and experimental work. However, the question about the exact
nature of the state generated by the radiation travelling through
the ensemble has only been raised very recently. Scully and
coworkers pointed out \cite{Scully2006}, that the dipole moments
of scatterers distributed along the incident beam's optical axis
are excited \emph{in phase} with beam's propagating phase front.
The resulting 'timed Dicke' state emits light predominantly into
forward direction, provided the ensemble size is large compared to
the radiation wavelength \cite{EPJD}.

This simple picture, however, only holds at low optical densities.
At high optical densities and spherically symmetric ensembles, the
refraction index of the scattering medium delays the propagation of the
pump beam and distorts its phase front. This distortion can have a huge
impact on any macroscopic observable of the system, such as the
angular distribution of the scattered radiation, the phase-front of the
transmitted beam or the force acting on the center-of-mass of the ensemble.

The aim of this paper is to calculate the correct state generated in an ensemble
of two-level systems, e.g.~an atomic cloud,
by interaction with a weak laser beam and how this state cooperatively
scatters the incident light.
Cooperative scattering by many atoms has been studied extensively
in the past, both classically and quantum-mechanically.
Classically, scattering at extended objects is described as Mie
scattering, showing resonances induced by the boundary conditions
that the target imposes to the incident light field \cite{Hulst}.
Quantum-mechanically, Dicke \cite{Dicke} has shown that, when
two-level atoms are confined inside a volume much smaller than a
radiation wavelength, the emission can be superradiant or
subradiant. How the classical and quantum pictures are linked has
been demonstrated at the example of a sample of
weakly excited atoms \cite{Prasad,Svi10}. More precisely, when a
single atom out of $N$ is excited, the Dicke symmetric state of
maximum cooperation radiates superradiantly, i.e. at a decay rate
proportional to $N$. Cooperative effects related to the
superradiant and directional emission by an extended ensemble of
atoms in a timed symmetric Dicke state have been
observed in the radiation pressure force acting on a large cloud
of atoms driven by a resonant radiation field. Depending on the detuning
of the incident radiation frequency from atomic resonance,
the radiation pressure force may be either
drastically reduced due to both increased forward scattering and a
reduced scattering cross section \cite{EPJD,Bienaime}, or even
enhanced if the cooperative Mie scattering dominates over
superradiance \cite{Bender}.

In this paper we revisit the scattering by $N$ atoms driven by a
constant uniform radiation field and emitting radiation into free
space. Our description assumes several approximations: 1) weak
excitation of the atomic ensemble (one atom out of $N$ is
excited); 2) Markov or 'rapid transit' approximation~\cite{Prasad}
(photon time of flight through the cloud much shorter than atomic
decay time); 3) atoms frozen (zero temperature) and motionless; 4)
neglected dipole-dipole interactions and collisions; 5) no
resonant interactions, i.e. no resonance fluorescence, Van der
Waals interactions etc. Approximation 3) excludes several
cooperative effects related to atomic recoil motion (e.g.
collective atom recoil lasing \cite{CARL,Slama} or matter wave
superradiance \cite{MWSR,Fallani}), and thus neglects stimulated
scattering processes along preferential directions, as for
instance end-fire modes in Bose-Einstein condensates \cite{Moore}
or optical cavity modes \cite{Bux}. Neglecting atomic interactions
(approximation 4) is justified assuming atomic distances much
larger than an optical wavelength. The yet interesting opposite
regime (atoms closer than an optical wavelength) requires the
solution of the atomic equations with an exponential interaction
kernel (see Eq.(\ref{eqbetaj})) and will be discussed in a future
publication.

We here determine the stationary solution for a spherical Gaussian
distribution, as well as relevant macroscopic
quantities such as medium polarization,
scattered radiation intensity and radiation pressure force. Our solution is based on a solution of the eigenvalue problem in the smooth density
approximation. For
large size samples, a continuous spectrum limit allows to
obtain explicit analytical expressions for such quantities,
expressed as a function of experimentally controllable parameters
such as frequency and power of the driving field, optical
thickness and size of the atomic sample.

The paper is organized as follow: in Sec.II the scattering
problem is expressed in terms of multi-particle coupled
differential equations. In Sec.III the continuous density
approximation is introduced and the solution for the atomic field
is  found in terms of the discrete eigenvalues of the interaction
operator. Then, some macroscopic quantities of importance are calculated in
Sec.IV. A continuous spectrum approximation for large atomic
clouds performed in Sec.V appears to be particularly suitable to
evaluate macroscopic quantities for arbitrarily large value of the
atomic density. Sec. VI revises the radiation pressure force
obtained assuming a symmetric timed Dicke atomic state \cite{EPJD} and,
comparing this state to the exact solution, we find that
phase shifts induced by the atomic cloud's refractive index are at
the origin of important corrections for the expectation values of
macroscopic quantities. Finally, numerical results and conclusions
are presented in Sec.VII.

\section{Cooperative scattering problem}

A system of two-level ($g$ and $e$) atoms with resonant frequency
$\omega_a$ and position $\mathbf{r}_j$, driven by an uniform laser
beam with electric field amplitude $E_0$, frequency $\omega_0$ and
wave vector $\mathbf{k}_0=(\omega_0/c)\mathbf{\hat e}_z$, is
described by the interaction Hamiltonian:
\begin{eqnarray}\label{H}
H&=&\frac{\hbar\Omega_0}{2}\sum_{j=1}^N\left[\hat\sigma_j
e^{i(\Delta_0 t- \mathbf{k}_0\cdot \mathbf{r}_j)}+\textrm{h.c.}\right]\nonumber\\
&+& \hbar\sum_{j=1}^N\sum_{\mathbf{k}}g_k\left(\hat\sigma_j
e^{-i\omega_a t} +\hat\sigma_j^\dagger e^{i\omega_a t}\right)
\left[\hat a_{\mathbf{k}}^\dagger e^{i(\omega_k t- \mathbf{k}\cdot
\mathbf{r}_j)}+\hat a_{\mathbf{k}} e^{-i(\omega_k t-
\mathbf{k}\cdot \mathbf{r}_j)}\right]
\end{eqnarray}
where $\Omega_0=d E_0/\hbar$ is the Rabi frequency of the incident
laser field, $\hat\sigma_j$ is the lowering operator for atom $j$,
$\hat a_{\mathbf{k}}$ is the photon annihilation operator and
$g_k=(d^2\omega_k/2\hbar\epsilon_0 V_{ph})^{1/2}$ is the
single-photon Rabi frequency, where $d$ is the electric-dipole
transition matrix element and $V_{ph}$ is the photon volume. A
special case is when a single photon is present in the mode
$\mathbf{k}$, as was extensively investigated in
Refs.\cite{FHM,Scully2006,Svi08}. The total system (atoms+photons)
is assumed to be in a state of the form \cite{Svi10}:
\begin{eqnarray}\label{state}
    |\Psi\rangle&=&\alpha(t)|g_1\dots g_N\rangle |0\rangle_{\mathbf{k}}+e^{-i\Delta_0 t}\sum_{j=1}^N
    \beta_j(t)|g_1\ldots e_j\ldots g_N\rangle|0\rangle_{\mathbf{k}}+ \sum_{\mathbf{k}}\gamma_{\mathbf{k}}(t)|g_1\dots g_N\rangle
    |1\rangle_{\mathbf{k}}\nonumber\\
    &+&\sum_{m,n=1}^N
    \epsilon_{m<n,\mathbf{k}}(t)|g_1\ldots e_m\ldots e_n\ldots
    g_N\rangle|1\rangle_{\mathbf{k}},
\end{eqnarray}
where $\Delta_0=\omega_0-\omega_a$. The first term corresponds to
the initial ground state without photons, the sum in the second
term are the states where a single atom has been excited by the
classical field. The third term corresponds to the atom returned
to the ground state having emitted a photon in the mode
$\mathbf{k}$, whereas the last one corresponds to the presence of
two excited atoms and one virtual photon with 'negative' energy.
It is due to the counter-rotating terms in the Hamiltonian
(\ref{H}) and disappears when the Rotating Wave Approximation
(RWA) is made.
 In the linear regime (i.e. $\alpha\approx 1$) and in the Markov
 approximation (valid if the decay time is larger than the
 photon time-of-flight through the atomic cloud), the problem reduces to the following differential
 equation \cite{Scully09},
 \begin{equation}\label{eqbetaj}
    \dot\beta_j=\left(i\Delta_0-\frac{\Gamma}{2}\right)\beta_j- i\frac{\Omega_0}{2}e^{i \mathbf{k}_0\cdot
    \mathbf{r}_j}+i\frac{\Gamma}{2}\sum_{m\neq j}
    \frac{\exp(ik_0|\mathbf{r}_j-\mathbf{r}_m|)}{k_0|\mathbf{r}_j-\mathbf{r}_m|}\beta_m,
\end{equation}
where $\Gamma=V_{ph}g_k^2 k_0^2/\pi c$ is the single-atom {\it
spontaneous} decay rate. The kernel in the last term of
Eq.(\ref{eqbetaj}) has a real component,  $-(\Gamma/2)\sum_{m\neq
j}[\sin(\rho_{jm})/\rho_{jm}]$ (where
$\rho_{jm}=k_0|\mathbf{r}_j-\mathbf{r}_m|$), describing the {\it
collective} atomic decay, and an imaginary component,
$i(\Gamma/2)\sum_{m\neq j}[\cos(\rho_{jm})/\rho_{jm}]$, describing
the collective Lamb shift due to short range interaction between
atoms, induced by the electromagnetic
field~\cite{Scully09,Scully10,Ralfie}. The latter becomes
significant when the number of atoms in a cubic optical
wavelength, $n\lambda^3$, is larger than unity, in which case the
contribution from the virtual processes described by the
counter-rotating terms in the Hamiltonian becomes relevant. Hence,
for sufficiently dilute system, such that $N\ll\sigma^3$ where
$\sigma=k_0\sigma_R$ and $\sigma_R$ is the cloud size, the
collective phase shift arising from the imaginary part of the
kernel in Eq.(\ref{eqbetaj}) can be disregarded~\cite{Svid09} and
the scattering problem reduces to
 \begin{equation}\label{eqbeta-sin}
    \dot\beta_j=\left(i\Delta_0-\frac{\Gamma}{2}\right)\beta_j- i\frac{\Omega_0}{2}e^{i \mathbf{k}_0\cdot
    \mathbf{r}_j}-\frac{\Gamma}{2}\sum_{m\neq j}
    \frac{\sin(k_0|\mathbf{r}_j-\mathbf{r}_m|)}{k_0|\mathbf{r}_j-\mathbf{r}_m|}\beta_m.
\end{equation}
with initial condition $\beta_j(0)=0$, for $j=1,\dots,N$. Notice
that Eq.(\ref{eqbetaj}) deduced in the quantum mechanical
description may be also obtained classically when the two-level
atoms are treated as weakly excited classical harmonic oscillators
\cite{Prasad,Svi10}. For this reason the solution of
Eq.(\ref{eqbetaj}) (or of the approximated version
(\ref{eqbeta-sin})) has a wider interest for the general problem
of collective radiation scattering.

As for the radiation field operator $\hat a_{\mathbf{k}}$, it
evolves according the Heisenberg equation
\begin{equation}\label{aH}
    \frac{d\hat
    a_{\mathbf{k}}}{dt}=\frac{1}{i\hbar}[\hat a_{\mathbf{k}},\hat
    H]=-ig_k e^{i(\omega_k-\omega_a) t}\sum_{m=1}^N \hat\sigma_m e^{-i\mathbf{k}\cdot
    \mathbf{r}_m},
\end{equation}
where the fast oscillating term proportional to
$\exp[i(\omega_k+\omega_a)t]$ has been neglected.

\section{Continuous density approximation}

In light scattering experiments, disorder (or granularity) plays a
role when the number of atoms projected onto a cross section
perpendicular to the incident beam is small enough so that a light
mode focused down to the diffraction limit (that is
$\sim\lambda^2$) would be able to resolve and count the atoms. In
other words, the stochastic fluctuations induced by the random
positions of the atoms can be neglected when the total number of
atoms $N$ is larger than the number of modes $\sim\sigma^2$ that
fit into the cloud's cross section, i.e. when the optical density
is $b_0=3N/\sigma^2\gg 1$. Under this hypothesis, the particles
can be described by a smooth density $n(\mathbf{r})$, and their
probability to be excited by a field $\beta(\mathbf{r},t)$, so
that Eq.(\ref{eqbeta-sin}) turns into
\begin{equation}\label{eqbeta-sin-cont}
    \frac{\partial}{\partial t}\beta(\mathbf{r},t)=\left(i\Delta_0-\frac{\Gamma}{2}\right)\beta(\mathbf{r},t)
    - i\frac{\Omega_0}{2}e^{i \mathbf{k}_0\cdot \mathbf{r}}-\frac{\Gamma}{2}\int \mbox{d}\mathbf{r}' n(\mathbf{r}')
    \frac{\sin(k_0|\mathbf{r}-\mathbf{r}'|)}{k_0|\mathbf{r}-\mathbf{r}'|}\beta(\mathbf{\mathbf{r}}',t).
\end{equation}
In what follows we will consider only spherically symmetric
distributions $n(r)$. Because of the linearity of
Eq.(\ref{eqbeta-sin-cont}), it is convenient to introduce an
eigenbasis of the coupling operator. The functions $j_n(r)
Y_{nm}(\theta,\phi)$, with $j_n$s the spherical Bessel functions
and $Y_{nm}(\theta,\phi)$s the spherical harmonics, appear as a
natural choice considering the following identity
\begin{equation}\label{sinc}
    \frac{\sin(k_0|\mathbf{r}-\mathbf{r}'|)}{k_0|\mathbf{r}-\mathbf{r}'|}=
    4\pi\sum_{n=0}^\infty\sum_{m=-n}^n j_n(k_0 r)Y_{nm}^*(\theta,\phi)Y_{nm}(\theta',\phi')j_n(k_0 r').
\end{equation}
In particular, the choice of the spherical harmonics guarantees the orthogonality of the basis, since
\begin{equation}
\int_0^{2\pi}\mbox{d}\phi\int_0^\pi \mbox{d}\theta\sin\theta
Y^*_{nm}(\theta,\phi)Y_{n'm'}(\theta,\phi)=\delta_{nn'}\delta_{mm'}\label{norm}.
\end{equation}
Therefore, assuming the following decomposition for the field,
\begin{equation}\label{betaexp}
    \beta(\mathbf{r},t)=\sum_{n=0}^\infty\sum_{m=-n}^n\alpha_{nm}(t)j_n(k_0
    r)Y_{nm}(\theta,\phi),
\end{equation}
the projection of Eq.(\ref{eqbeta-sin-cont}) along the eigenmodes leads to
\begin{eqnarray}
    &&\left[
    \dot\alpha_{n m}-i\Delta_0\alpha_{nm}+\frac{\Gamma}{2}(1+\lambda_n)\alpha_{n,m}\right]
    j_{n}(k_0 r)=-\frac{i}{2}\Omega_{nm},
    \label{Ebeta3}
\end{eqnarray}
where $\lambda_n$ are the eigenvalues associated to modes $n$
\begin{equation}\label{eigen}
    \lambda_{n}=4\pi\int_0^\infty r^2 n(r)j_{n}^2(k_0
    r) \mbox{d}r
\end{equation}
whereas $\Omega_{nm}$ corresponds to the projection of the incident wave on mode $(n,m)$
\begin{equation}\label{Omega}
    \Omega_{n m}=\Omega_0
    \int_0^{2\pi}\mbox{d}\phi\int_0^\pi d\theta\sin\theta\,
    Y^*_{n m}(\theta,\phi)\,
    e^{ik_0r\cos\theta}=2\Omega_0\delta_{m 0}\sqrt{\pi(2n+1)}i^n j_{n}(k_0 r).
\end{equation}
Assuming the cloud is initially unexcited, i.e.
$\alpha_{mn}(0)=0$, only spherically symmetric components with
$m=0$ will grow so that, defining
$\alpha_n(t)\equiv\alpha_{n0}(t)$, Eq.(\ref{Ebeta3}) reduces to
\begin{eqnarray}
    \dot\alpha_{n}-\left[i\Delta_0-\frac{\Gamma}{2}(1+\lambda_n)\right]\alpha_{n}
    =-i^{n+1}\sqrt{\pi(2n+1)}\Omega_0
    \label{alphadot}
\end{eqnarray}
Eq.(\ref{alphadot}) straightforwardly integrates and, inserted in Eq.(\ref{betaexp}), leads to the following expression for the excitation field
\begin{eqnarray}
    \beta(r,\theta,t)&=&\frac{\Omega_0}{\Gamma}\sum_{n=0}^\infty\frac{i^{n}(2n+1)j_n(k_0 r)P_{n}(\cos\theta)}{2\delta+i(1+\lambda_n)}
    \left[1-e^{i\Delta_0 t}e^{-(\Gamma/2)(1+\lambda_n) t}\right],\label{betafin1}
\end{eqnarray}
where the scaled detuning $\delta=\Delta_0/\Gamma$ was introduced.
Hence, each mode $n$ relaxes toward the steady-state with a
characteristic time $\tau_n=1/\Gamma(1+\lambda_n)$: the first
modes relax very quickly since $\lambda_n$ is proportional to $N$,
yet for the highest modes, $\tau_n\sim \Gamma^{-1}$, even if their
macroscopic contribution is usually small. Eventually, for times
much longer than the single-atom decay time $\Gamma^{-1}$, the
field tends toward a stationary state fully characterized by the
spectrum
\begin{eqnarray}
    \beta_s(r,\theta)&=&\frac{\Omega_0}{\Gamma}\sum_{n=0}^\infty\frac{i^{n}(2n+1)}{2\delta+i(1+\lambda_n)}j_n(k_0 r)P_{n}(\cos\theta).\label{beta:ss}
    \end{eqnarray}
Notice that the set of eigenvalues (\ref{eigen}) is complete since
from the identity $\sum_{n\geq0} (2n+1)j_n^2(z)=1$, it follows that
\begin{equation}\label{ident3}
\sum_{n=0}^\infty (2n+1)\lambda_n=4\pi\int_0^\infty r^2 n(r) \mbox{d}r=N,
\end{equation}
which corresponds to the trace of the coupling operator.

\section{Macroscopic quantities}

The description of the field $\beta_s(r,\theta)$ in terms of
spectrum also provides expressions for any macroscopic quantities,
the most relevant of which are calculated here below. These
formulae will be specialized to Gaussian clouds in the subsequent
section.

\subsection{Average amplitude and  probability of excitation}

The average 'phased' probability of the timed Dicke state \cite{Scully09}
and the excitation probability are respectively
\begin{eqnarray}
  \langle\beta_s e^{-i\mathbf{k}_0\cdot \mathbf{r}}\rangle &=& \frac{2\pi}{N}
  \int_0^\pi \mbox{d}\theta\sin\theta\int_0^\infty \mbox{d}r r^2 n(r)\beta_s(r,\theta)
  e^{-ik_0r\cos\theta}\\
  \langle|\beta_s|^2\rangle &=& \frac{2\pi}{N}\int_0^\pi \mbox{d}\theta\sin\theta\int_0^\infty \mbox{d}r r^2
  n(r)|\beta_s(r,\theta)|^2.
\end{eqnarray}
Inserting Eq.(\ref{beta:ss}) and using the identities
\begin{equation}\label{ident}
    \int_{-1}^1 \mbox{d}x P_n(x)e^{i\alpha
    x}=2i^nj_n(\alpha)\quad,\quad
    \int_{-1}^1
    dx\,P_m(x)P_n(x)=\frac{2}{2n+1}\delta_{m n},
\end{equation}
we obtain
\begin{eqnarray}
    \langle\beta_s e^{-i\mathbf{k}_0\cdot \mathbf{r}}\rangle &=&\frac{\Omega_0}{\Gamma N}
    \sum_{n=0}^\infty \frac{(2n+1)\lambda_n}{2\delta+i(1+\lambda_n)}
    \label{ave:beta}\\
    \langle|\beta_s|^2\rangle &=& \frac{\Omega_0^2}{\Gamma^2 N}\sum_{n=0}^\infty
    \frac{(2n+1)\lambda_n}{4\delta^2+(1+\lambda_n)^2}.
    \label{ave:beta2}
\end{eqnarray}

\subsection{Scattered field}

The electric field radiated by the excited atoms reads, in the
smooth density limit (see Appendix A),
\begin{equation}\label{Es2}
    {\cal E}_S(r,\theta,\phi,t)=-\frac{dk_0^2}{4\pi\epsilon_0 r}e^{ik_0(r-ct)}
    \int
    \mbox{d}\mathbf{r}'n(\mathbf{r}')\beta(\mathbf{r}',t)
    e^{-i\mathbf{k}_s\cdot \mathbf{r}'}.
\end{equation}
Using the stationary solution (\ref{beta:ss}) for a spherically
symmetric distribution and using the integral
\begin{eqnarray}\label{int:Pro}
\int_0^\pi
    \mbox{d}\theta'\sin\theta'P_n(\cos\theta')J_0(k_0r'\sin\theta\sin\theta')
    e^{-ik_0r'\cos\theta\cos\theta'}=2 i^{-n} j_n(k_0
    r')P_n(\cos\theta)
\end{eqnarray}
we obtain
\begin{equation}\label{Int2}
    {\cal E}_S(r,\theta)=-\left(\frac{E_0}{4\pi k_0 r}\right)  e^{ik_0(r-ct)}
    \sum_{n=0}^\infty \frac{(2n+1)\lambda_n}{2\delta+i(1+\lambda_n)}
    P_n(\cos\theta),
\end{equation}
where we used the relation $\Gamma=d^2
k_0^3/(2\pi\hbar\epsilon_0)$. In the forward direction
($\theta=0$) the scattered field is proportional to the 'phased'
probability amplitude (\ref{ave:beta}). The expression
(\ref{Int2}) provides the angular distribution of the scattered
radiation field. We can also calculate the scattered intensity as
\begin{equation}\label{Int:sca}
    I_S(r,\theta,\phi)=c\epsilon_0\langle\hat E_S^\dagger \hat E_S\rangle=c\epsilon_0
    \left(\frac{dk_0^2}{4\pi\epsilon_0 r}\right)^2\left[
    \sum_{j=1}^N|\beta_j|^2+\sum_{j\neq
    m}\beta_m^*\beta_je^{-i\mathbf{k}_s\cdot(\mathbf{r}_j-\mathbf{r}_m)}\right]
\end{equation}
Passing to the continuous limit and using Eqs.(\ref{ave:beta2}),
(\ref{Es2}) and (\ref{Int2}), we obtain:
\begin{equation}\label{Int:sca2}
    I_S(r,\theta)=
     \frac{I_0}{(4\pi k_0 r)^2}\left[
    \sum_{n=0}^\infty \frac{(2n+1)\lambda_n}{4\delta^2+(1+\lambda_n)^2}
    +\left|\sum_{n=0}^\infty \frac{(2n+1)\lambda_n}{2\delta+i(1+\lambda_n)}
    P_n(\cos\theta)
    \right|^2\right]
\end{equation}
where $I_0=c\epsilon_0 E_0^2$. The scattered intensity is the sum
of the incoherent contribution, proportional to $N$ (since
$\lambda_n\propto N$, see Eq.(\ref{eigen})) and isotropic, and the
superradiant contribution, proportional to $N^2$ and directed (for
extended clouds) mainly in the forward direction. Integrating over
the solid angle, the total scattered power is
\begin{equation}\label{P:sca}
    P_S=2\pi r^2\int_0^\pi d\theta\sin\theta I_S(r,\theta)=
    \left(\frac{I_0}{4\pi k_0^2 }\right) \sum_{n=0}^\infty
    \frac{(2n+1)\lambda_n(1+\lambda_n)}{4\delta^2+(1+\lambda_n)^2}
\end{equation}
where we used the second of the identities (\ref{ident}).

\subsection{Radiation pressure force}

The radiation force operator acting on the $j${th} atom is
calculated from Eq.(\ref{H}) as
$\hat{\mathbf{F}}_j=-\nabla_{\mathbf{r}_j}\hat
H=\hat{\mathbf{F}}_{aj}+\hat{\mathbf{F}}_{ej}$ where \cite{EPJD}
\begin{eqnarray}
    \hat{\mathbf{F}}_{aj}&=& i\hbar \mathbf{k}_0\frac{\Omega_0}{2}
    \left\{\hat\sigma_{j} e^{i(\Delta_0 t-\mathbf{k}_0\cdot \mathbf{r}_j)}-
    \textrm{h.c.}\right\}\label{Force-abs}\\
   \hat{\mathbf{F}}_{ej}&=& i\hbar\sum_{\mathbf{k}} \mathbf{k}g_{k}
    \left\{\hat a_{\mathbf{k}}^\dagger \hat\sigma_{j}
    e^{i(\omega_k-\omega_a)t-i\mathbf{k}\cdot\mathbf{r}_j}-
    \hat\sigma_{j}^\dagger \hat a_{\mathbf{k}}
    e^{-i(\omega_k-\omega_a)t+i\mathbf{k}\cdot\mathbf{r}_j}\right\}
    \label{Force-emi}
\end{eqnarray}
are the forces  due respectively to the absorption and emission
processes. In Eq.(\ref{Force-emi}) we have neglected the
counter-rotating terms proportional to $\exp[\pm
i(\omega_k+\omega_a)t]$. We are here interested in the average
radiation force $\hat{\mathbf{F}}=(1/N)\sum_j\hat{ \mathbf{F}}_j$,
which stands for the force acting on the center-of-mass of the
atomic cloud along the direction of the incident field,
$\mathbf{k}_0=k_0 \mathbf{\hat e}_z$. This average force is
relatively easy to measure by time-of flight techniques in cold
atomic clouds released, for instance, from magneto-optical traps
(MOTs) and has recently revealed cooperative effects in the
scattering by extended atomic samples \cite{Bienaime,Bender}. It
may provide a convenient measurement (aside from the scattered
radiation) of the effects that cooperative scattering imprints on
the atoms. The average absorption force along the $z$-axis,
resulting from the recoil received upon absorption of a photon
from the incident laser, is
\begin{eqnarray}
 \hat F_a &=& =
 \frac{i}{2N}\hbar k_0\Omega_0\sum_{j=1}^N\left[\hat\sigma_j e^{i\Delta_0 t-i \mathbf{k}_0\cdot
 \mathbf{r}_j}-\textrm{h.c.}\right].
 \label{Fa}
\end{eqnarray}
The second contribution $\hat{\mathbf{F}}_e=(1/N)\sum_j\hat
{\mathbf{F}}_{ej}$ results from the emission of a photon into any
direction $\mathbf{k}$. Inserting $\hat a_{\mathbf{k}}$ from
Eq.(\ref{aH}) into Eq.(\ref{Force-emi}) and approximating the sum
over the modes $\mathbf{k}$ by an integral, it is possible to
obtain,  in a way similar as done for the radiation field operator
$\hat E_S$ of Eq.(\ref{Es}),  the following expression for the
average emission force along the $z$-axis \cite{EPJD}:
\begin{eqnarray}
   \hat F_e &=& -\frac{\hbar k_0\Gamma}{8\pi N}\int_0^{2\pi}\mbox{d}\phi\int_0^\pi
  \mbox{d}\theta\sin\theta\cos\theta\sum_{j,m=1}^N\left[
    e^{-i\mathbf{k}\cdot(\mathbf{r}_j-\mathbf{r}_m)}
    \hat\sigma_{m}^\dagger\hat\sigma_{j}+\textrm{h.c.}\right].
  \label{Fez}
\end{eqnarray}
Evaluating their expectation values on the state (\ref{state})
(neglecting virtual photon contributions), the emission force in
the discrete model is
\begin{eqnarray}\label{Force:emis}
    \langle F_{e}\rangle&=& -\frac{\hbar k_0\Gamma}{8\pi N}\int_0^{2\pi}\mbox{d}\phi\int_0^\pi
  \mbox{d}\theta\sin\theta\cos\theta\sum_{j,m=1}^N\left[\beta_{j}\beta_{m}^*
    e^{-i\mathbf{k}\cdot(\mathbf{r}_j-\mathbf{r}_m)}
    +\textrm{c.c.}\right]\nonumber\\
    &=& i\frac{\hbar k_0\Gamma}{2N}
    \sum_{j,m=1}^N
    \frac{(z_j-z_m)}{|\mathbf{r}_j-\mathbf{r}_m|}j_1(k_0|\mathbf{r}_j-\mathbf{r}_m|)(\beta_j\beta_m^*-\textrm{c.c.}).
\end{eqnarray}
where  we used the identity (\ref{id:j1}) and $j_1(z)$ is the
first order spherical Bessel function. Then, passing to the
continuous distribution limit, Eqs.(\ref{Fa}) and (\ref{Fez}) are
approximated by
\begin{eqnarray}
 \langle\hat F_a\rangle &=& -\hbar k_0\Omega_0 \textrm{Im}\langle\beta e^{-i \mathbf{k}_0\cdot \mathbf{r}}\rangle
 \label{Fa:cont} \\
 \langle\hat F_e\rangle &=& -\hbar k_0\frac{\Gamma}{8\pi N}\int_0^{2\pi}\mbox{d}\phi\int_0^\pi
  \mbox{d}\theta\sin\theta\cos\theta\int d\mathbf{r} n(r)\int
  \mbox{d}\mathbf{r}'n(r')\left[\beta(\mathbf{r})\beta^*(\mathbf{r}')
  e^{-i\mathbf{k}\cdot(\mathbf{r}-\mathbf{r}')}+\textrm{c.c.}\right].
  \label{Fez:cont}
\end{eqnarray}
The absorption stationary force is readily obtained from
Eq.(\ref{ave:beta}),
\begin{eqnarray}
 \langle\hat F_a\rangle &=& \hbar k_0\frac{\Omega_0^2}{\Gamma N}
\sum_{n=0}^\infty
\frac{ (2n+1)\lambda_n(1+\lambda_n)}{4\delta^2+(1+\lambda_n)^2}
\label{Fa:cont2}
\end{eqnarray}
whereas a longer calculation, reported in Appendix B, yields the
emission stationary force,
\begin{eqnarray}\label{Force:fur:1}
    \langle F_{e}\rangle
    &=& -\hbar
    k_0\frac{2\Omega_0^2}{\Gamma N}\sum_{n=0}^\infty
    \frac{(n+1)\lambda_n\lambda_{n+1}[4\delta^2+(1+\lambda_n)(1+\lambda_{n+1})]}
    {[4\delta^2+(1+\lambda_n)^2][4\delta^2+(1+\lambda_{n+1})^2]}.
\end{eqnarray}
\begin{figure}
$\begin{array}{cc}
\includegraphics[width=8cm]{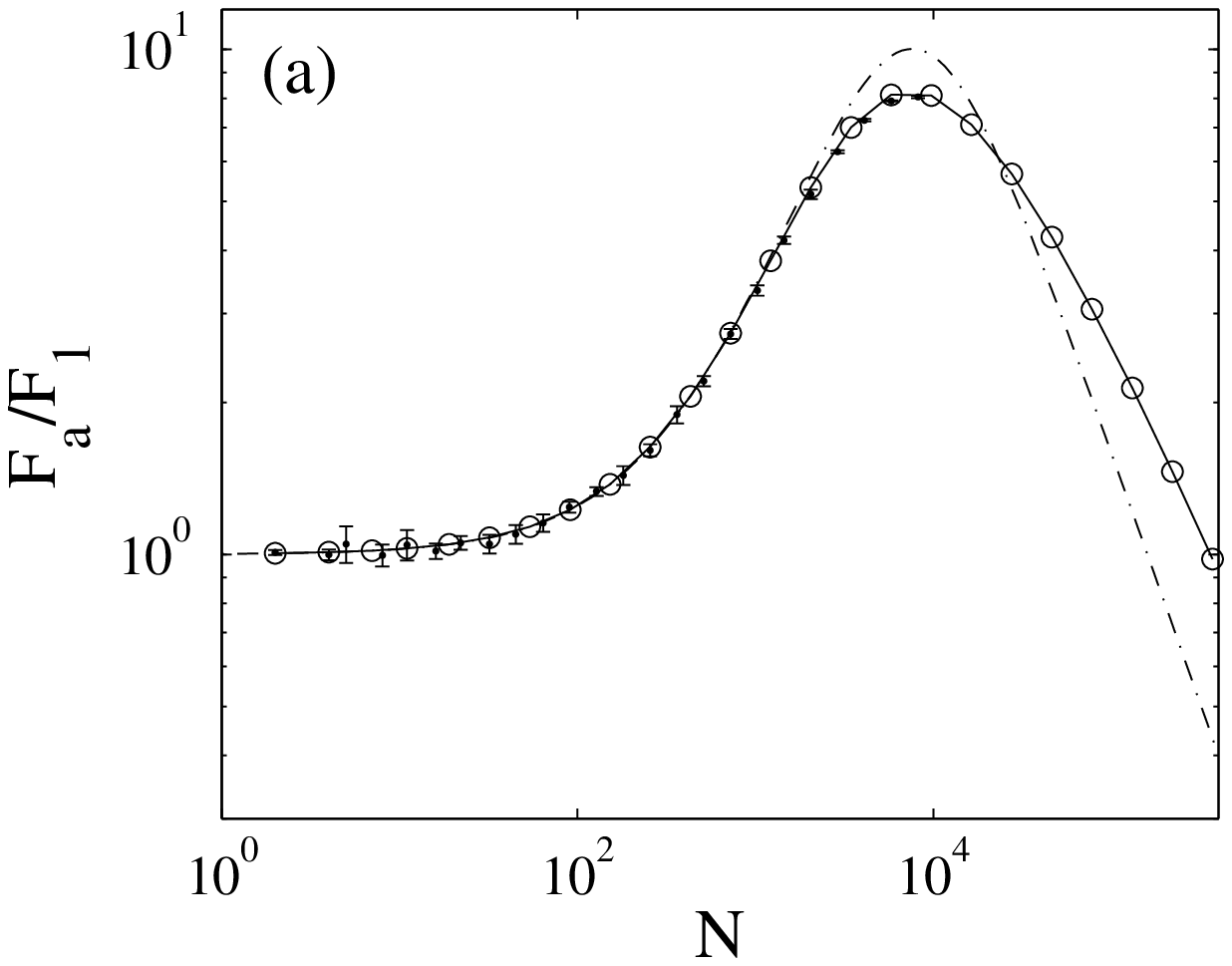}&\includegraphics[width=8cm]{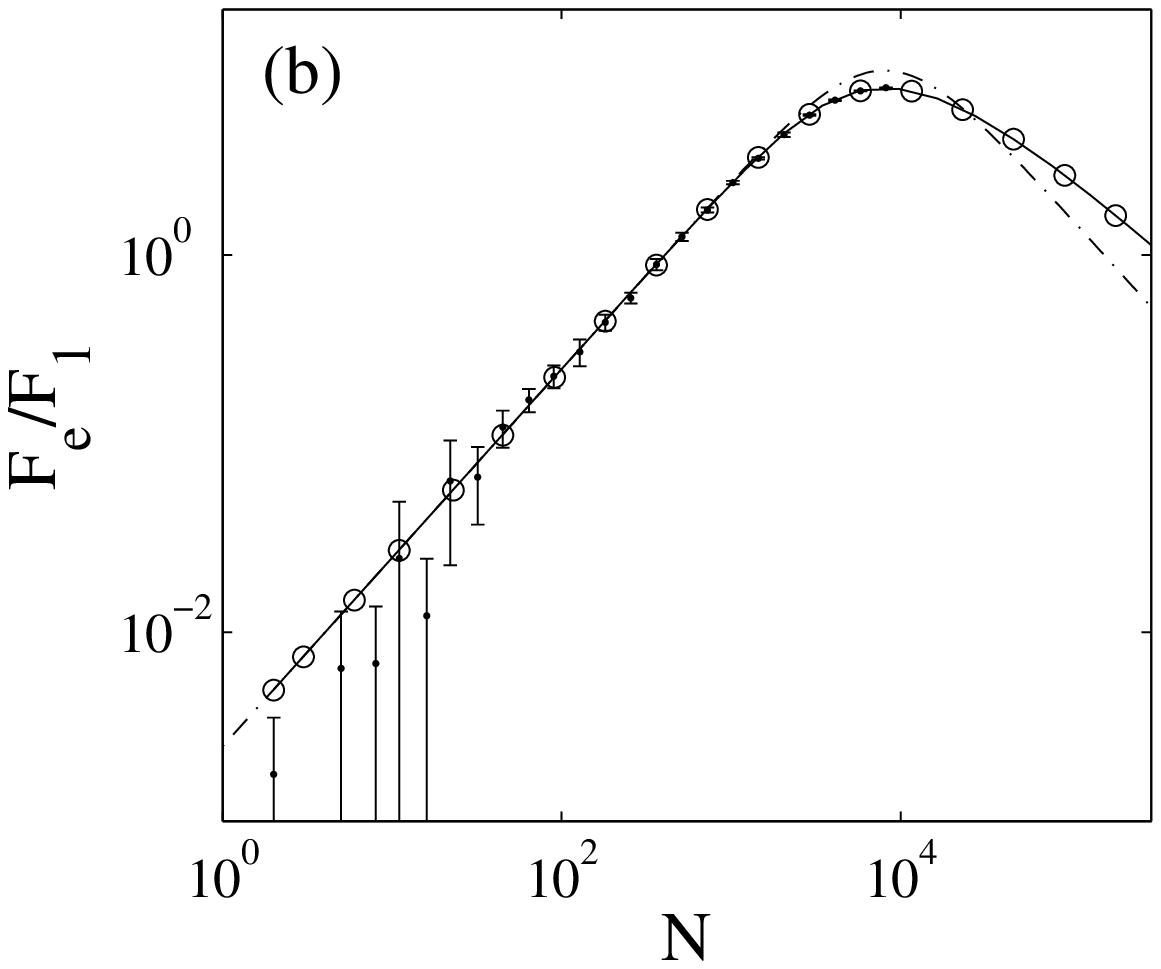}
\\ \includegraphics[width=8cm]{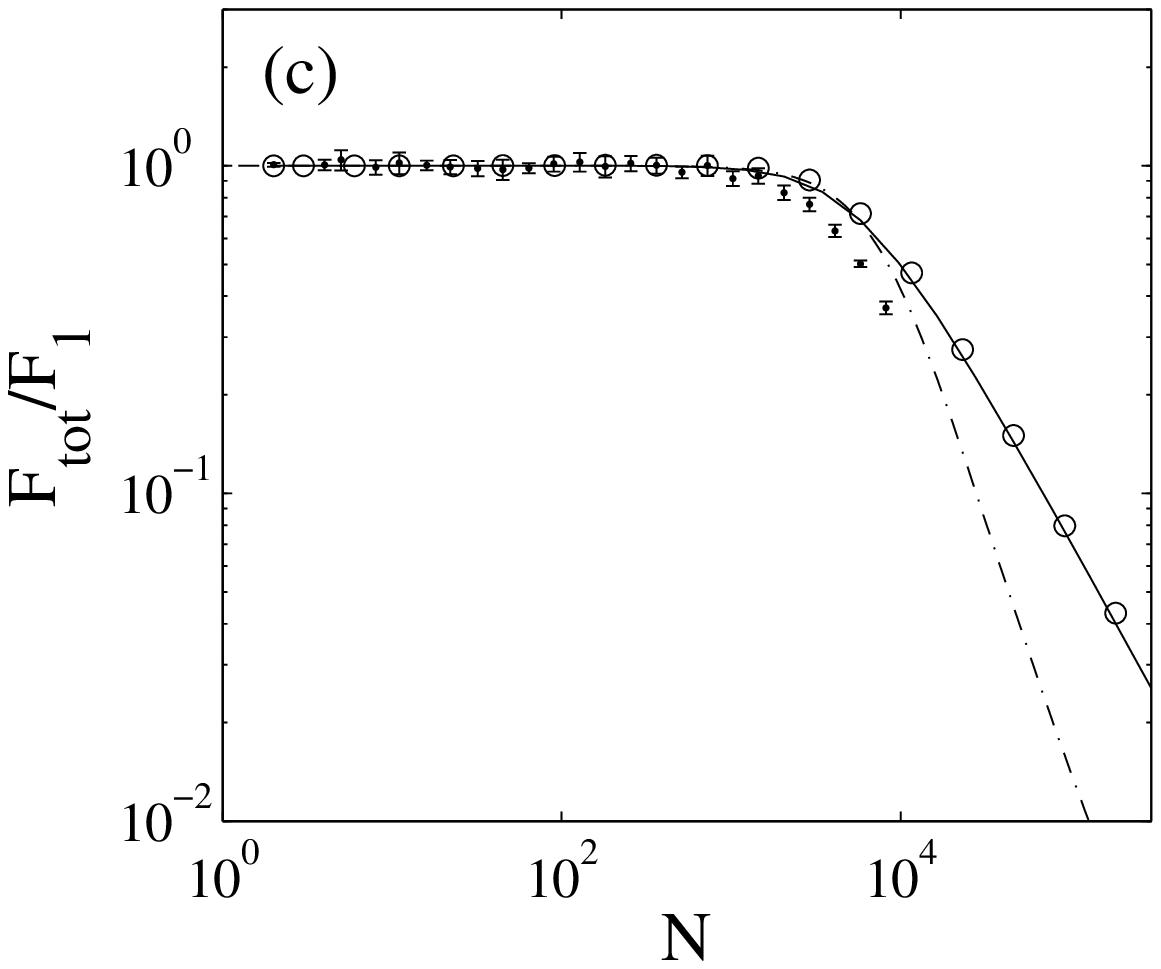}&\includegraphics[width=8cm]{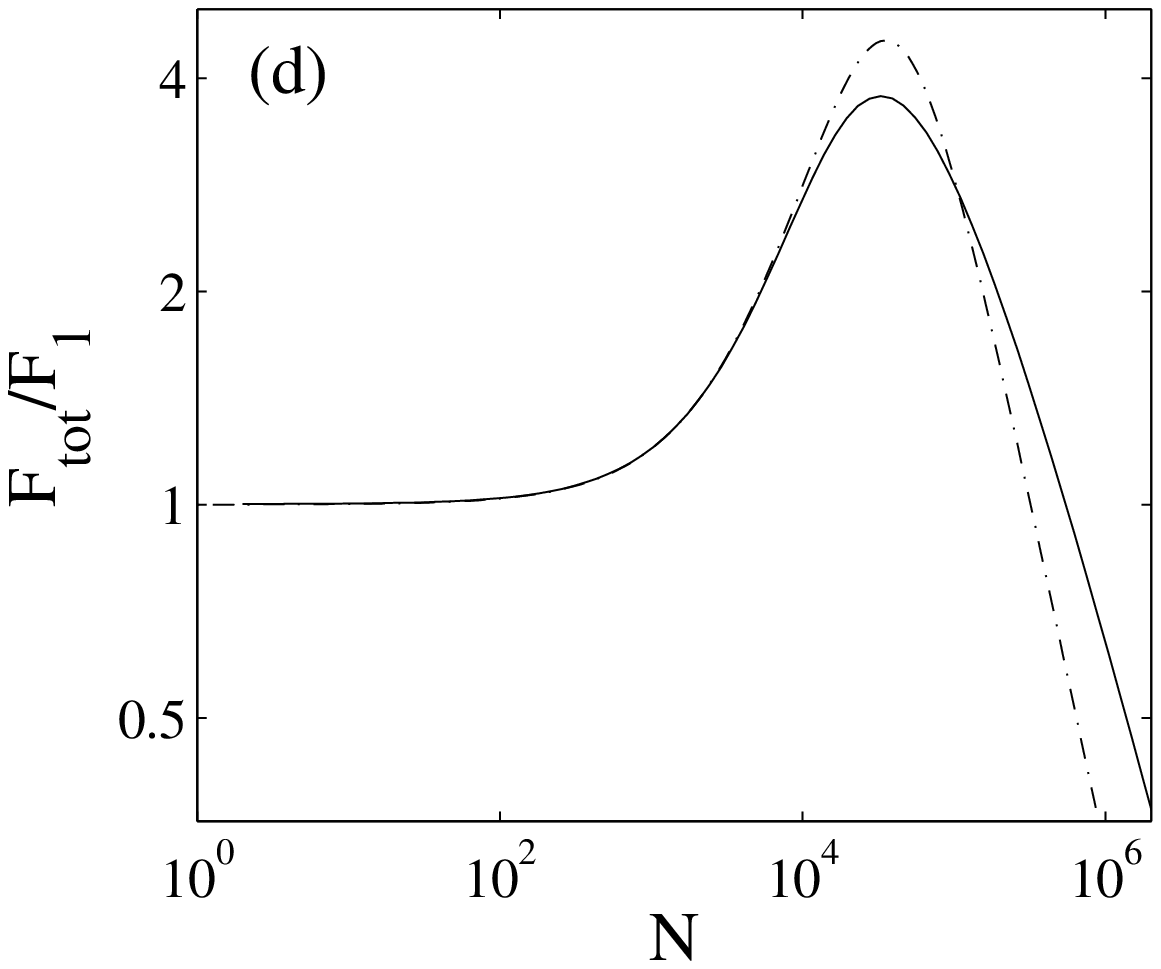}
\end{array}$
\caption{(a) Absorption, (b) emission and (c) total forces vs $N$
for $\delta=10$ and a Gaussian cloud with $\sigma=8$. (d) Total
force for $\sigma=5$ and $\delta=200$. Forces are relative to the
single-atom force $F_1$. The plain curves refer to the series
(\ref{Fa:cont2}), (\ref{Force:fur:1}) and their sum, the circles
to the analytical expressions (\ref{ave:beta:sol2}),
(\ref{Force:fur:1}) and their sum, the dots to $N$-body
simulations (see Eq.\ref{eqbeta-sin}) and the dash-dotted lines to
the STD state. The error bars correspond to the standard deviation
of the observables over $8$ realizations.\label{fig:forces}}
\end{figure}
As expected, the absorption force pushes the atomic cloud in the
direction of the driving field, whereas the emission force is
oppositely directed and is proportional to $N$. Both forces depend
on $N$ also through the collective decay rate
$\Gamma(1+\lambda_n)$. We observe that the absorption force
(\ref{Fa:cont2}) is linked to the scattered power (\ref{P:sca}) by
the relation
\begin{equation}\label{Qa}
N\langle\hat F_a\rangle=4\pi\left(\frac{P_S}{c}\right),
\end{equation}
i.e. the absorption force is proportional to the scattered power
per atom, $P_S/N$.

Fig.\ref{fig:forces} compares the (a) absorption, (b) emission and
(c) total forces vs. $N$ for $\delta=10$ and a Gaussian density
profile with $\sigma=8$ (where $\sigma=k_0\sigma_R$), calculated
using the series (\ref{Fa:cont2}) and (\ref{Force:fur:1})
(circles) and $N$-body simulations (see Eq.\ref{eqbeta-sin})
(dots). The eigenvalues $\lambda_n$ for the Gaussian density
profile are given by Eq.(\ref{eigen:gauss}). The forces are
reported as a ratio between the cooperative force and the
single-atom force, $F_1=\hbar
k_0\Gamma\Omega_0^2/(4\Delta_0^2+\Gamma^2)$. The error bars
correspond to the standard deviation of the observables over $8$
realizations. Fig.\ref{fig:forces}d shows the total force vs. $N$
for a different choice of parameters, $\sigma=5$ and $\delta=200$,
for which the force exhibits a maximum as a function of
$N$~\cite{Bender}. We observe a good agreement between the
analytical solution (\ref{Fa:cont2}) and the $N$-body simulations
at large $N$, when effects due to the discreteness of the system
become negligible.

We recall that our expressions have been obtained in the
continuous density approximation, i.e. assuming a sample with
large optical thickness, i.e. $N\gg\sigma^2$, but sufficiently
dilute to neglect the collective Lamb shift, i.e. with a small
number of atoms in a cubic optical wavelength volume, $N\ll
\sigma^3$. These two conditions imply that the cloud size should
be much larger than the optical wavelength, $\sigma\gg 1$.
Nevertheless, our results remain valid also in the limit
$n(r)\rightarrow 0$ where, in particular, the standard radiation
pressure force, $F_1$, is recovered. However, at low densities,
fluctuations will strongly affect individual measurements of the
radiation pressure force and in this regime our expressions
represent only the average expectation values.

\section{Large Gaussian clouds\label{sec:lclouds}}

Large clouds ($\sigma\gg1$) behaves fundamentally differently from small clouds ($\sigma\lesssim 1$), as can be deduced from their spectrum.
Let us for example consider the case of Gaussian clouds, with density $n(r)=(N/(2\pi)^{3/2}\sigma_R^3)\exp(-r^2/2\sigma_R^2)$.
The spectrum then reads
\begin{equation}
\lambda_n=N\sqrt{\frac{2}{\pi}}\int_0^\infty \rho^2 e^{-\rho^2/2}j_n^2(\sigma \rho) \mbox{d}\rho=N\sqrt{\frac{\pi}{2\sigma^2}}e^{-\sigma^2}
I_{n+1/2}(\sigma^2),
\label{eigen:gauss}
\end{equation}
where $\sigma=k_0\sigma_R$ is the scaled size of the cloud, and
$I_n(x)$ the $n$-th modified Bessel function. Recently the
spectrum $\lambda_n$ for an uniform spherical cloud has been
calculated by A.A. Svidzinsky et al. \cite{Svi08,Svi10}, also for
the exponential kernel of Eq.(\ref{eqbetaj}). However, a
Gaussian distribution is certainly more realistic for experiments
with cold dense atomic ensembles. Generally, the
spectrum of small clouds ($\sigma\le 1$) is composed of a few
significant eigenmodes, whereas for $\sigma$ large, all the
eigenmodes for $n<\sigma$ are significant and the spectrum can be
treated as a continuum. In particular, in the latter case the
$\lambda_{n}$ can be approximated, for $n<\sigma$, by
$\lambda_n\sim (N/2\sigma^2)\exp[-(n+1/2)^2/2\sigma^2]$ (see
e.g.~\cite{Abra}). Switching to a continuous treatment of the
spectrum, we define $\eta=n+1/2$ and get
\begin{eqnarray}
\lambda_\eta=\frac{N}{2\sigma^2}e^{-\eta^2/(2\sigma^2)}.\label{eq:leta}
\\ \sum_{n=0}^\infty (2n+1)\rightarrow 2\int_0^\infty \eta \mbox{d}\eta.
\end{eqnarray}
Remark that using these definitions, the completeness condition
(\ref{ident3}) is still preserved. The continuous spectrum limit
allows for the evaluation of the sums in Eqs.(\ref{ave:beta}) and
(\ref{ave:beta2}) as continuous integrals
\begin{eqnarray}
    \langle\beta_{s}e^{-i\mathbf{k}_0\cdot \mathbf{r}}\rangle&=&\frac{2\Omega_0}{\Gamma N}\int_{0}^\infty
    \frac{ \lambda_\eta\eta\mbox{d}\eta}{2\delta+i(1+\lambda_\eta)}
    =\frac{\Omega_0}{\Gamma} \frac{6}{b_0}\int_{0}^{b_0/6}
    \frac{\mbox{d}x}{2\delta+i(1+x)}
    \label{ave:beta:asy}\\
    \langle|\beta_{s}|^2\rangle &=& \frac{2\Omega_0^2}{\Gamma^2 N}\int_{0}^\infty
    \frac{\lambda_\eta\eta \mbox{d}\eta}{4\delta^2+(1+\lambda_\eta)^2}
    =\left(\frac{\Omega_0}{\Gamma}\right)^2\frac{6}{b_0}\int_{0}^{b_0/6}
    \frac{\mbox{d}x}{4\delta^2+(1+x)^2}
    \label{ave:beta2:asy}
\end{eqnarray}
where we have set $x=(b_0/6)\exp(-\eta^2/2\sigma^2)$, with
$b_0=3N/\sigma^2$ the optical thickness. The above expressions
integrate as
\begin{eqnarray}
    \langle\beta_{s}e^{-i\mathbf{k}_0\cdot \mathbf{r}}\rangle&=&\left(\frac{\Omega_0}{\Gamma}\right)
    \frac{3}{b_0}\left\{
    2\arctan\left[\frac{b_0}{3}\frac{\delta}{1+4\delta^2+b_0/6}\right]-i
    \ln\left[1+\frac{b_0}{3}\frac{1+b_0/12}{1+4\delta^2}\right]
    \right\}
    \label{ave:beta:sol1}\\
    \langle|\beta_{s}|^2\rangle &=& \left(\frac{\Omega_0}{\Gamma}\right)^2\frac{3}{\delta b_0}
    \arctan\left[\frac{\delta b_0/3}{1+4\delta^2+b_0/6}\right].
    \label{ave:beta2:sol2}
\end{eqnarray}
These formulae highlight the prominent role of the parameters
$b_0$ and $\delta$ in the high-density limit. In a similar way, we
calculate the total scattered power as:
\begin{equation}\label{Itot}
    P_S=\frac{I_0 N}{4\pi k_0^2}\int_0^1 dx\frac{1+(b_0/6)x}{4\delta^2+[1+(b_0/6)x]^2}=\frac{I_0\sigma_R^2}{4\pi}
    \ln\left[1+\frac{b_0}{3}\frac{1+b_0/12}{1+4\delta^2}\right]
\end{equation}
As expected, for small optical thickness ($b_0\ll 1$) the
scattered power is incoherent, $P_S\approx [I_0 N/(4\pi
k_0^2)]/(4\delta^2+1)$. However, for large optical thickness it
shows a logarithmic dependence on $N$. The superradiant character
of the radiation is visible only observing the scattered intensity
in the forward direction (see the second term of
Eq.(\ref{Int:sca2})), but not in the total scattered power.

The absorption force is deduced from Eqs.(\ref{Fa:cont}) and
(\ref{ave:beta:sol1}) as
\begin{equation}
\langle\hat F_a\rangle = \hbar k_0 \frac{3\Omega_0^2}{b_0\Gamma}\ln\left[1+\frac{b_0}{3}\frac{1+b_0/12}{1+4\delta^2}\right].\label{ave:beta:sol2}
\end{equation}

The emission force can be written, in the continuous spectrum
approximation in the integral form
\begin{eqnarray}\label{Force:cont:ss1}
    \langle\hat F_{e}\rangle
    &=& -\hbar k_0\frac{\Omega_0^2 b_0}{6\Gamma}e^{-1/(4\sigma^2)}\int_{1/\sigma}^\infty
    dy
\frac{y
e^{-y^2}\left\{4\delta^2+\left[1+(b_0/6)e^{-(y+1/2\sigma)^2/2}\right]\left[1+(b_0/6)e^{-(y-1/2\sigma)^2/2}\right]\right\}}
    {\left\{4\delta^2+\left[1+(b_0/6)e^{-(y+1/2\sigma)^2/2}\right]^2\right\}
    \left\{4\delta^2+\left[1+(b_0/6)e^{-(y-1/2\sigma)^2/2}\right]^2\right\}},
\end{eqnarray}
where we have set $y=(\eta+1)/\sigma$. As can be observed in Fig.\ref{fig:forces},
the continuous-spectrum approximation gives excellent results compared to the full series (\ref{Fa:cont2}) and (\ref{Force:fur:1}). In the limit
$\sigma\rightarrow\infty$ and finite $b_0$,
Eq.(\ref{Force:cont:ss1}) would lead to
\begin{equation}\label{Fe:approx}
\langle\hat F_e\rangle \approx -\langle\hat F_a\rangle+\hbar
k_0\Gamma \langle|\beta_s|^2\rangle,
\end{equation}
which has a transparent interpretation: for a very large cloud,
the atoms scatter radiation to forward direction and the recoil
received by the atoms upon emission cancels out with the recoil
received by absorbing a photon from the driving field. The net
force remaining after the substraction  is the non collective
contribution to the emission. The net force is equal to the photon
momentum $\hbar k_0$ times the emission rate $\Gamma
\langle|\beta_s|^2\rangle$. This emission rate depends indirectly
on $N$ and $\sigma$ through the enhanced superradiant decay,
$\Gamma (N/4\sigma^2)$, which decreases the emission rate when the
optical thickness increases. A more accurate expression of the
emission force valid for large but finite cloud size would require
the exact evaluation of the integral in Eq.(\ref{Force:cont:ss1}).

\section{Symmetric Timed Dicke state\label{sec:STD}}

A particular ansatz used by Scully and
coworkers~\cite{Scully2006,Svi08,Svi10} is the symmetric timed
Dicke (STD) state, given by
\begin{equation}\label{STD}
    \beta(\mathbf{r},t)=\beta_{TD}(t)e^{i\mathbf{k}_0\cdot \mathbf{r}}.
\end{equation}
After integration over space of Eq.(\ref{eqbeta-sin-cont}), one obtains the following evolution equation
\begin{equation}\label{eqbeta-TD}
    \frac{d\beta_{TD}(t)}{dt}=\left[i\Delta_0-\frac{\Gamma}{2}(1+Ns_\infty)\right]\beta_{TD}(t)
    - i\frac{\Omega_0}{2},
\end{equation}
where $s_\infty$ is the integrated structure factor of the cloud defined as
\begin{equation}\label{SN}
    s_\infty=\frac{1}{N^2}\int n(\mathbf{r}) \mbox{d}\mathbf{r}\int \mbox{d}\mathbf{r}' n(\mathbf{r}')
    \frac{\sin(k_0|\mathbf{r}-\mathbf{r}'|)}{k_0|\mathbf{r}-\mathbf{r}'|}e^{-i \mathbf{k}_0\cdot
    (\mathbf{r}-\mathbf{r}')}
    =\frac{1}{4\pi}\int_0^{2\pi}\mbox{d}\phi\int_0^\pi \mbox{d}\theta\sin\theta
    |\langle e^{i(\mathbf{k}_0-\mathbf{k})\cdot
    \mathbf{r}}\rangle|^2.
\end{equation}
This ansatz is of particular interest since it allows to evidence
e.g. the superradiant nature of the decay of such state when the
pump is turned off \cite{Scully09}. As for its steady-state, it
reads \cite{EPJD,Bienaime}
\begin{equation}\label{STDS}
    \beta_{TD}=\frac{\Omega_0}{\Gamma}\frac{1}{2\delta+i(1+N s_{\infty})},
\end{equation}
and for a large cloud with Gaussian distribution, since $s_\infty\approx
1/4\sigma^2$, we get
\begin{equation}\label{STDS:G}
    \beta_{TD}=\frac{\Omega_0}{\Gamma}\frac{1}{2\delta+i(1+b_0/12)}.
\end{equation}
Thus, the STD solution (\ref{STDS:G}) approximates the exact
result (\ref{ave:beta2:sol2}) only for $b_0\ll 12$,
\begin{equation}\label{STDS:G2}
    |\beta_{TD}|^2\approx\left(\frac{\Omega_0}{\Gamma}\right)^2\frac{1}{4\delta^2+(1+b_0/6)}.
\end{equation}
\begin{figure}
$\begin{array}{cc}
\includegraphics[width=8cm]{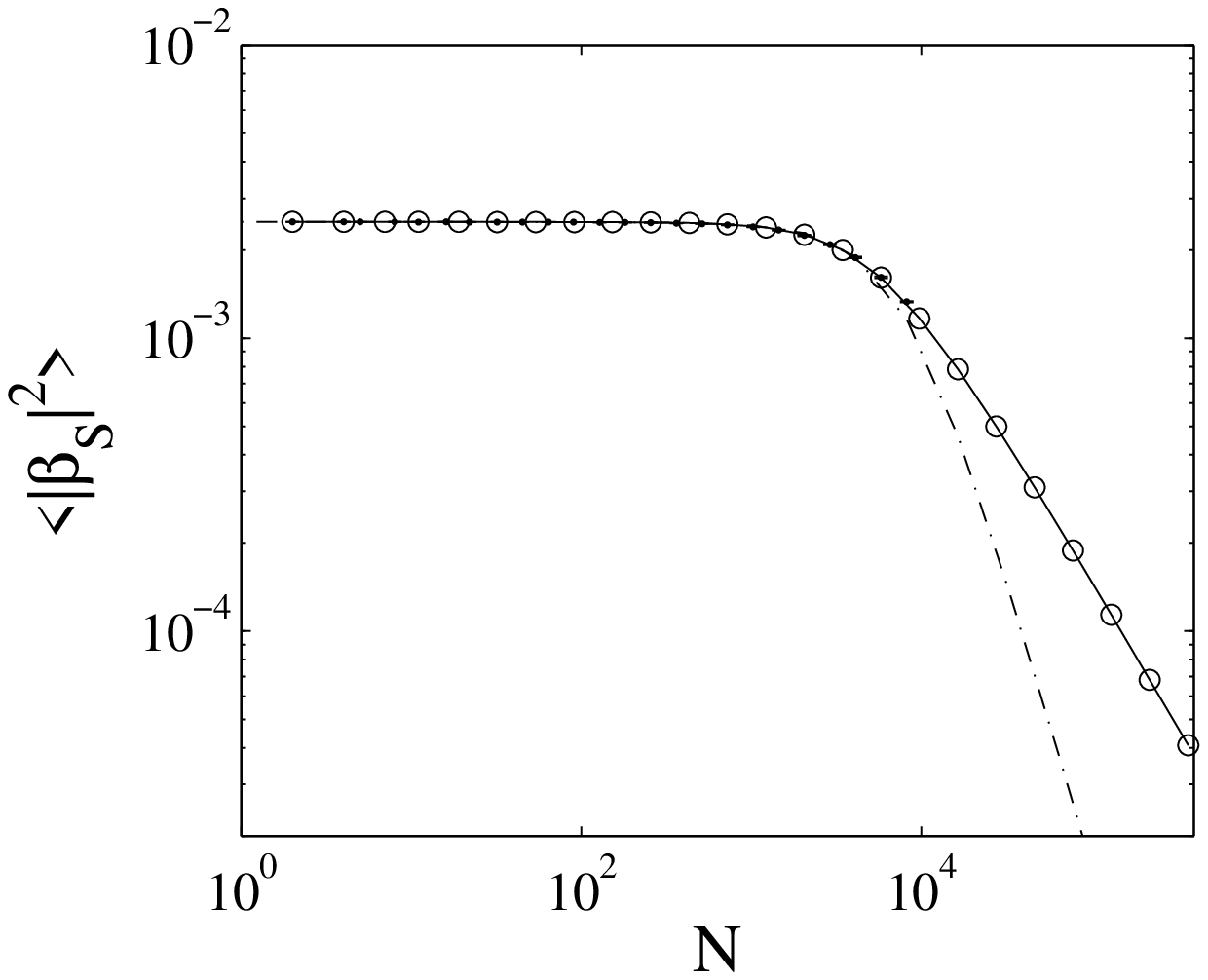}&\includegraphics[width=8cm]{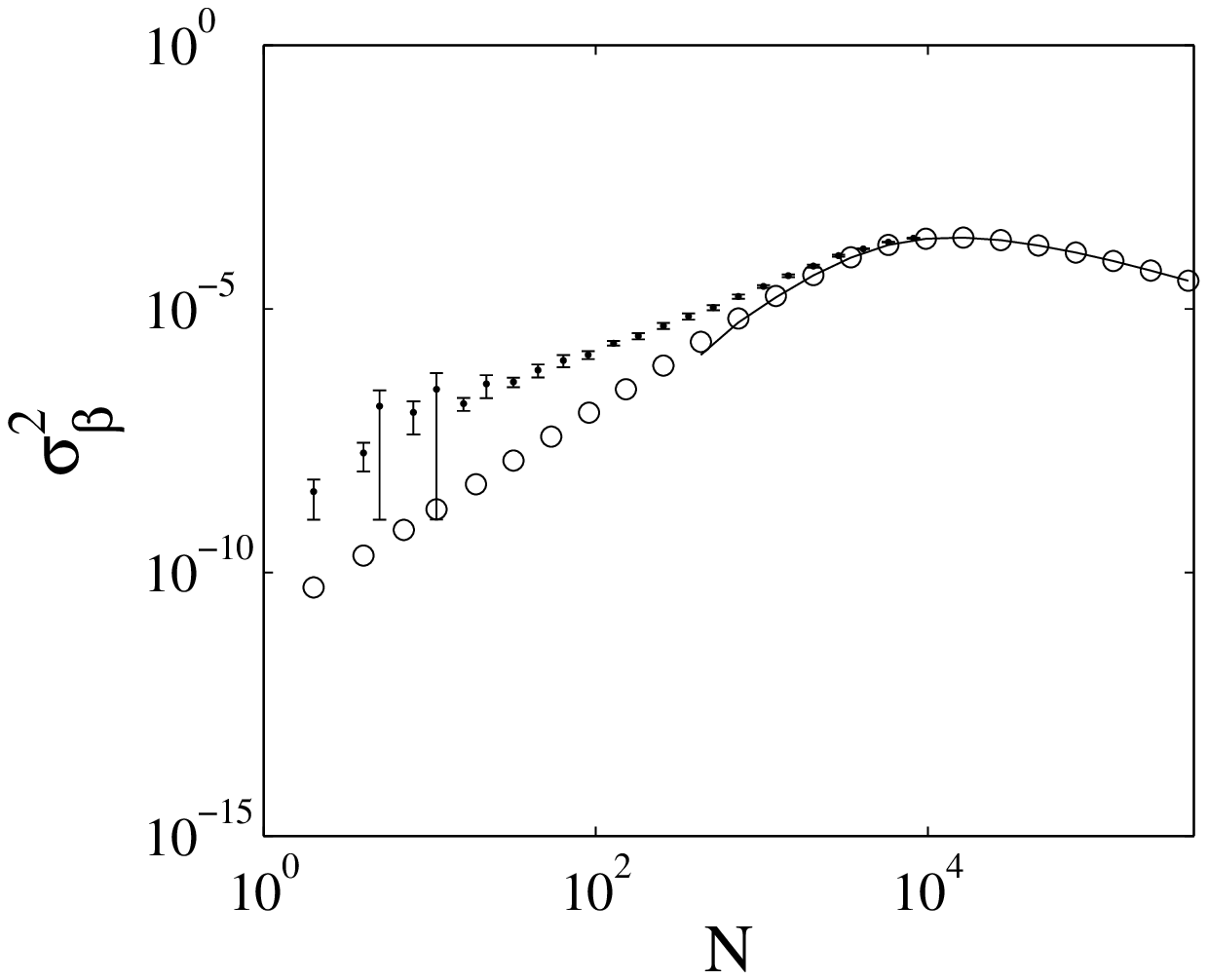}
\end{array}$
\caption{Average excitation probability
$\langle|\beta_s|^2\rangle$ (left) and variance
$\sigma_\beta^2=\langle|\beta_s|^2\rangle-|\langle \beta_s e^{-i
k_0.\mathbf{r}} \rangle|^2$ (right) vs $N$. The plain curves refer
to the analytical expressions (\ref{ave:beta:sol1}) and
(\ref{ave:beta2:sol2}), the circles to series expressions
(\ref{ave:beta}) and (\ref{ave:beta2}), the dots to $N$-body
simulations (see Eq.\ref{eqbeta-sin}) and the dash-dotted lines to
the STD state. The error bars correspond to the standard deviation
of the observables over $8$ realizations. Note that for the STD
state, $\sigma_\beta=0$; for too small $N$, the approximations
(\ref{ave:beta:sol1}) and (\ref{ave:beta2:sol2}) provide
inconsistent results, that is negative $\sigma_{\beta}$.
Simulations realized for $\sigma=10$ and $\delta=10$.}
\label{fig:betam}
\end{figure}
Fig.\ref{fig:betam} shows the average excitation probability
$\langle|\beta_s|^2\rangle$ (left)  and its variance,
$\sigma_\beta^2=\langle|\beta_s|^2\rangle-|\langle \beta_s e^{-i
k_0.\mathbf{r}} \rangle|^2$ vs. $N$ for $\sigma=10$ and
$\delta=10$. The plain curves refer to the analytical expressions
(\ref{ave:beta:sol1}) and (\ref{ave:beta2:sol2}), the circles to
series expressions (\ref{ave:beta}) and (\ref{ave:beta2}), the
dots to $N$-body simulations (see Eq.\ref{eqbeta-sin}) and the
dash-dotted lines to the STD state (\ref{STD}). The error bars
correspond to the standard deviation of the observables over
different realizations. Note that for the STD state,
$\sigma_\beta=0$; for too small $N$, the approximations
(\ref{ave:beta:sol1}) and (\ref{ave:beta2:sol2}) provide
inconsistent results, that is negative $\sigma_{\beta}$. We
observe excellent agreement between the series and the analytical
solutions, and a consistent reduction of the excitation
probability decrease vs $N$ with respect to the TDS prediction
(dashed line in fig.\ref{fig:betam}, left). Also the fluctuations
obtained from the $N$-body simulations converge for large values
of $N$ toward the results obtained in the continuous density
approximation (fig.\ref{fig:betam}, left), showing the presence of
a shot-noise contribution for small $N$.

The radiation force for the STD ansatz (\ref{STD}) is
\cite{EPJD,Bienaime,Bender}
\begin{equation}\label{TDForce}
    \langle\hat F\rangle=\langle\hat F_a\rangle+\langle\hat
    F_e\rangle=\hbar k_0\left[-\Omega_0 \textrm{Im}(\beta_{TD})-\Gamma |\beta_{TD}|^2
    N f_{\infty}\right]
\end{equation}
where
\begin{equation}\label{fN}
    f_\infty=\frac{1}{4\pi}\int_0^{2\pi}\mbox{d}\phi\int_0^\pi
\mbox{d}\theta\sin\theta\cos\theta
    |\langle e^{i(\mathbf{k}_0-\mathbf{k})\cdot
    \mathbf{r}}\rangle|^2.
\end{equation}
Since for a spherically Gaussian distribution
$N(s_{\infty}-f_{\infty})\approx N/(8\sigma^4)=b_0/(24\sigma^2)$,
from Eq.(\ref{STDS}) the stationary radiation force is
\cite{EPJD,Bienaime,Bender}
\begin{eqnarray}\label{TDForce1}
    \langle\hat F\rangle&=&\hbar k_0\Gamma\left(\frac{\Omega_0}{\Gamma}\right)^2\frac{1+N(s_\infty-f_\infty)}{4\delta^2+(1+Ns_{\infty})^2}
    \\ &\approx& \hbar
    k_0\Gamma\left(\frac{\Omega_0}{\Gamma}\right)^2\frac{1+b_0/24\sigma^2}{4\delta^2+(1+b_0/12)^2}.
\end{eqnarray}
As can be observed in Fig.\ref{fig:forces}, the STD state yields a
good agreement with the full-spectrum approach and the $N$-body
simulations only for small values of the optical thickness.

Finally, for that state (\ref{STD}) the scattered radiation
electric field, Eq.(\ref{Es}), and intensity, (\ref{Int:sca}),
become
\begin{equation}\label{Es:STD}
    {\cal E}_S(r,\theta,\phi,t)=\frac{dk_0^2}{4\pi\epsilon_0 r}e^{ik_0(r-ct)}
    \beta_{TD}(t)\langle e^{i(\mathbf{k}_0-\mathbf{k})\cdot \mathbf{r}}\rangle
\end{equation}
and
\begin{equation}\label{Is:STD}
    I_S(r,\theta,\phi)=c\epsilon_0
    \left(\frac{dk_0^2}{4\pi\epsilon_0 r}\right)^2|\beta_{TD}|^2\left[
    N+N^2|\langle e^{i(\mathbf{k}_0-\mathbf{k})\cdot \mathbf{r}}\rangle|^2\right].
\end{equation}
In particular, for a Gaussian distribution, $\langle
\exp[i(\mathbf{k}_0-\mathbf{k})\cdot
\mathbf{r}]\rangle=\exp[-\sigma^2(1-\cos\theta)/2]$.

\begin{figure}
$\begin{array}{ccc}
\includegraphics[width=6cm]{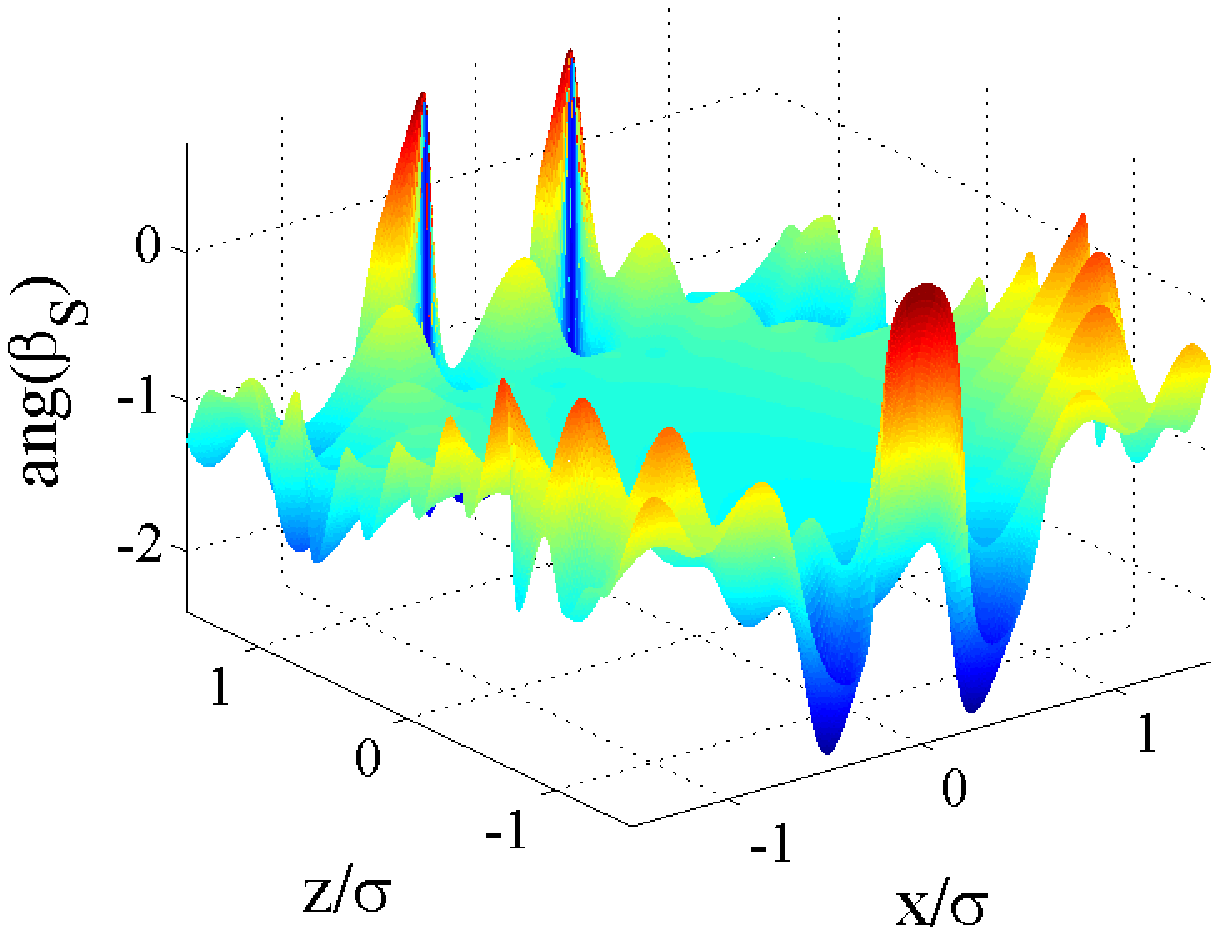}&\includegraphics[width=6cm]{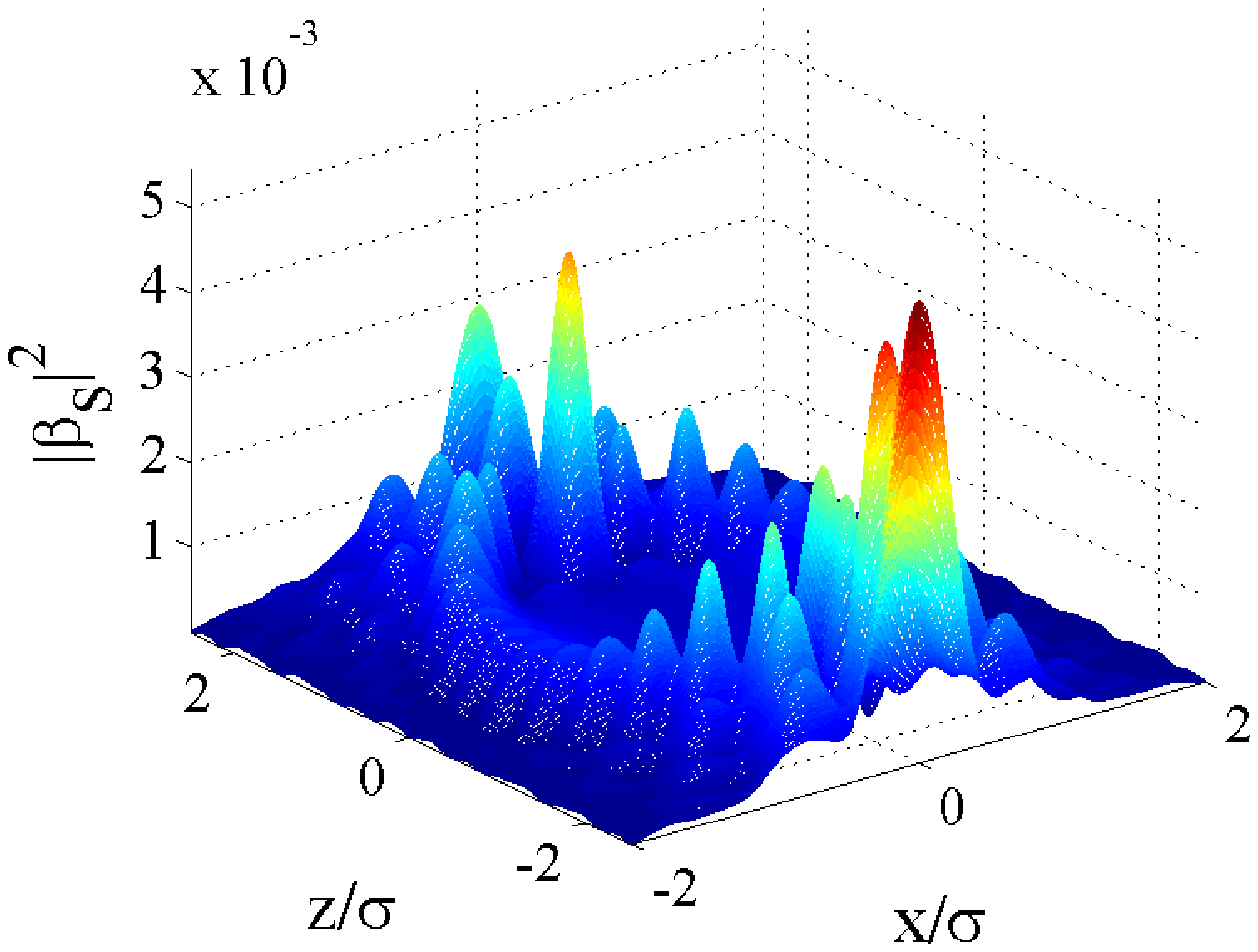}&\includegraphics[width=6cm]{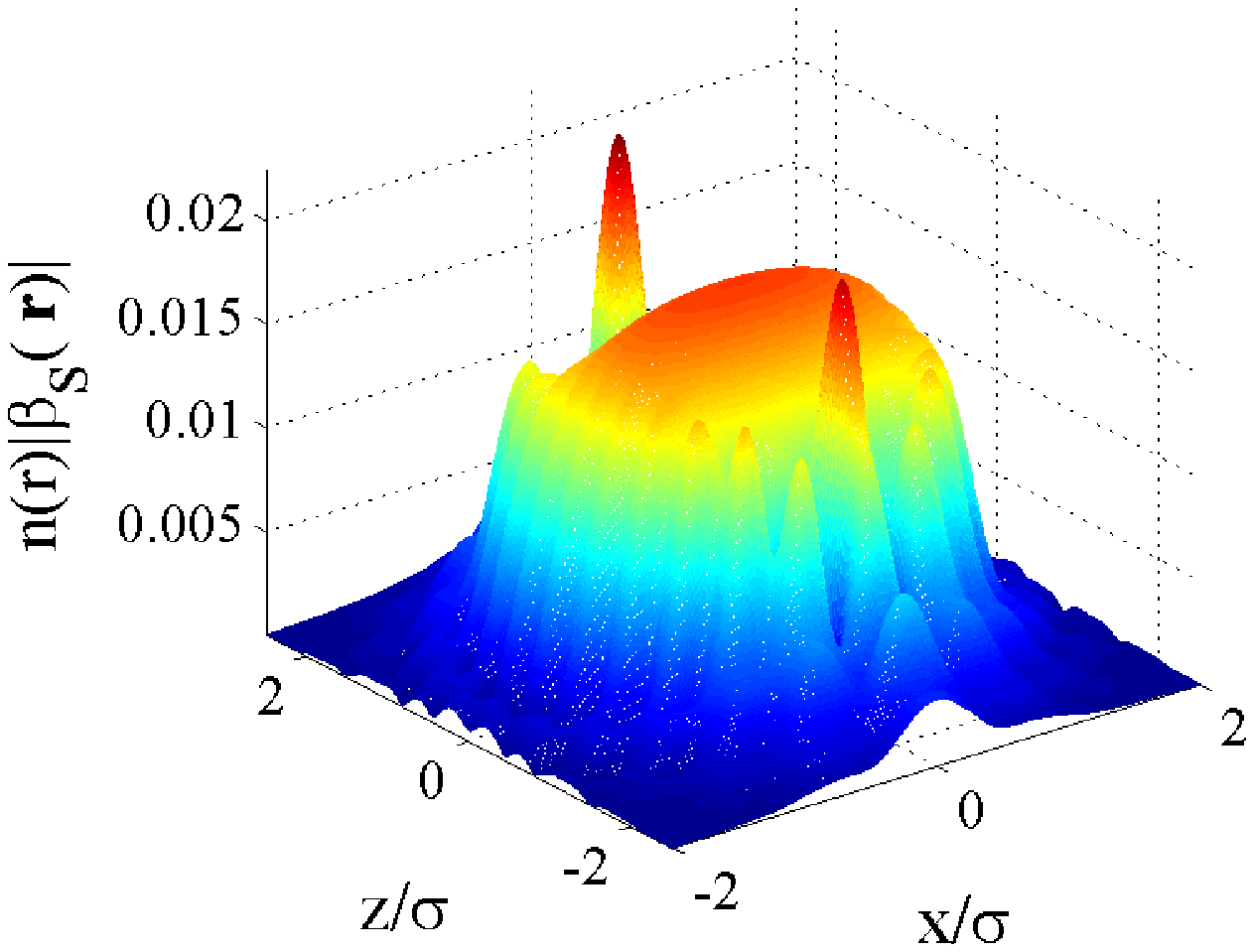}
\end{array}$
\caption{Phase of the atomic radiation $ang(\beta_s)$ (left),
Excitation probability $|\beta_s|^2$ of the atoms in the $(x,z)$
plane ($y=0$) (center) and contribution to the radiation of the
electric field in the same plane, which corresponds to the level
of excitation of the atoms weighted by their local density
(right). Simulations realized using the analytic expressions
(\ref{beta:ss}), for $N=10^4$, $\sigma=10$ and $\delta=10$.}
\label{fig:beta}
\end{figure}

\begin{figure}
$\begin{array}{cc}
\includegraphics[width=6cm]{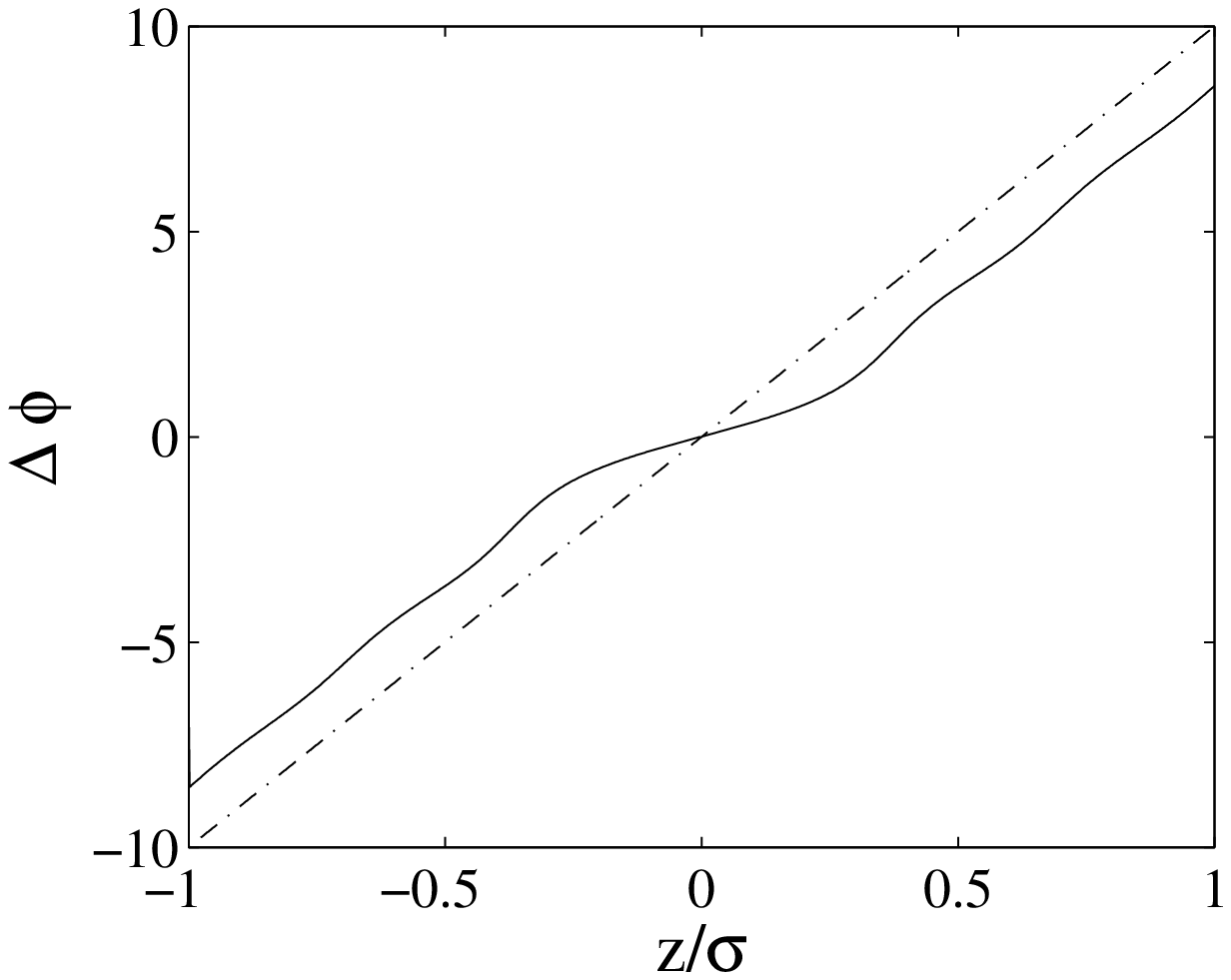}&\includegraphics[width=6cm]{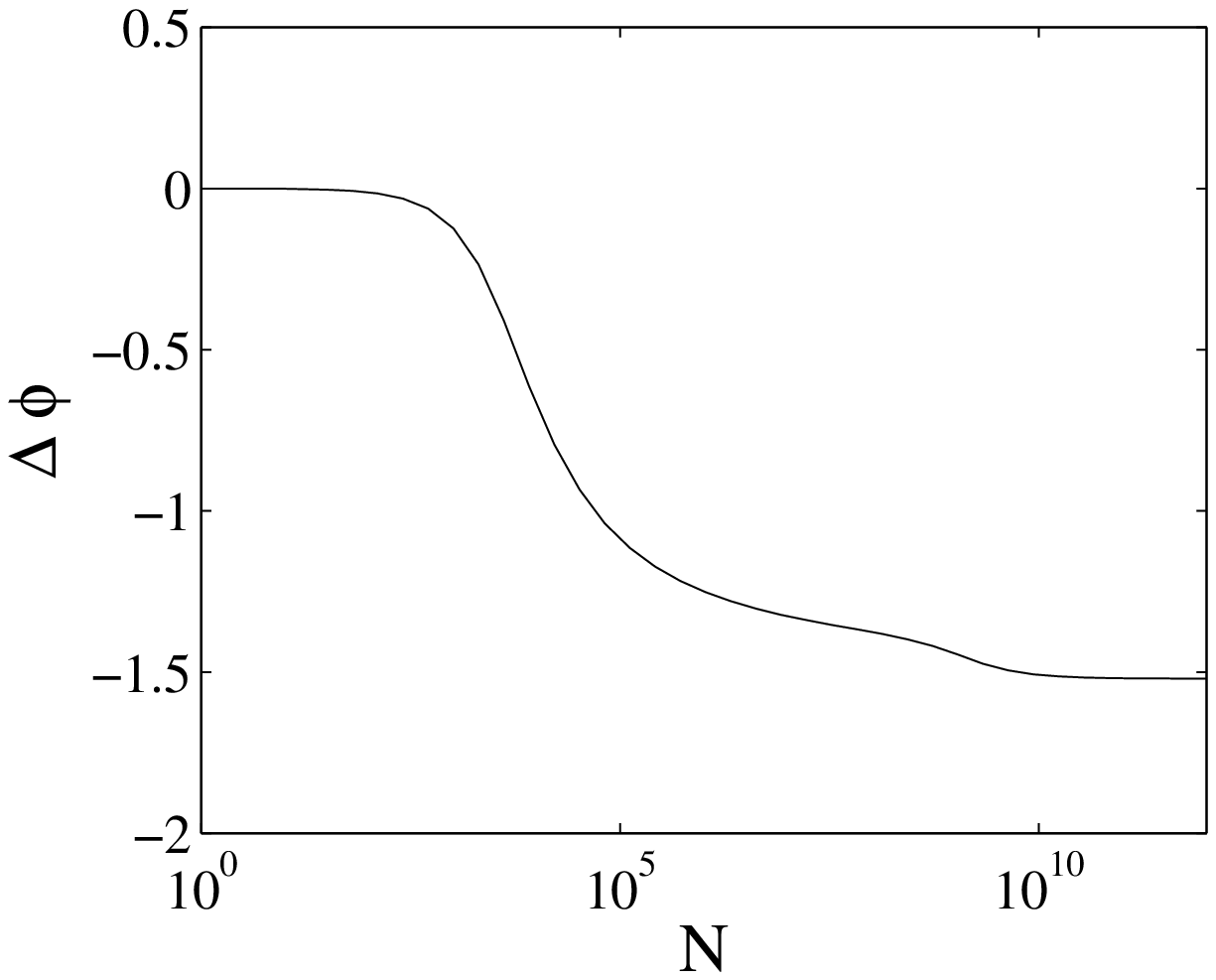}
\end{array}$
\caption{Left: Phase shift along the optical axis for the exact
solution (\ref{beta:ss}) (plain line) and a STD state (\ref{STD})
(dash-dotted line). Right: Phase shift of the pump beam after
transmission through the cloud as a function of $N$ evaluated
along the axis , $x=y=0$, at $k_0z=20\sigma$, for $N=10^4$,
$\sigma=10$ and $\delta=10$.} \label{fig:phaseshift}
\end{figure}

Fig.\ref{fig:beta} shows the phase of the excitation amplitude
$\beta_s$ (left) and the excitation probability, $|\beta_s|^2$,
(center) in the $(x,z)$ plane ($y=0$), calculated from the exact
solution (\ref{beta:ss}), for $N=10^4$, $\sigma=10$ and
$\delta=10$. Fig.\ref{fig:beta} (right) shows the contribution to
the electric field radiation in the same plane,  weighted by the
local atomic density. The simplification of the STD, as compared
to the exact solution (\ref{beta:ss}), resides in the assumption
that all atoms are equally excited and oscillate in phase.
According to the exact calculation, the atomic dipoles appear to
be in phase only in the core of the cloud (see Fig.\ref{fig:beta},
left), but this phase profile has strong distortion away from it.
This phenomenon is all the more important as the atoms are much
more excited in the peripheral region than in the core (see
Fig.\ref{fig:beta}, center). In particular, even when this
excitation probability is weighted by the particle density, two
areas at the cloud entrance and exit contribute significantly to
the radiation electric field (\ref{Int2}) (see peaked structures
in Fig.\ref{fig:beta}, right). For a timed Dicke state, both the
phase profile and average excitation remain flat throughout the
cloud.
From a macroscopic point of view, the STD state neglects phase
shifts imprinted into the pump beam by the cloud's reflective
index. This can be seen in Fig.~\ref{fig:phaseshift}(a), which
compares the phase of the STD state,
$\beta_{TD}\exp(i\mathbf{k}_0\cdot \mathbf{r})$, (linear curve, no
phase shift) and $\beta_s(\mathbf{r})$ (additional phase-shift)
along the optical axis across the cloud.
Fig.~\ref{fig:phaseshift}(b) shows the pump beam phase shift after
transmission through the atomic cloud as a function of atom
number. This phase shift is at the origin of the deviation between
the radiation pressure forces calculated for the STD state and the
exact solution.
The pump beam phase shift leads to a reduction of the absorption
and the emission forces. This can be understood as destructive
interference of forward radiation emitted from different atoms,
located at the same plane $z=z_0$ but different $x$ or $y$.

\begin{figure}
$\begin{array}{cc}
\includegraphics[width=6cm]{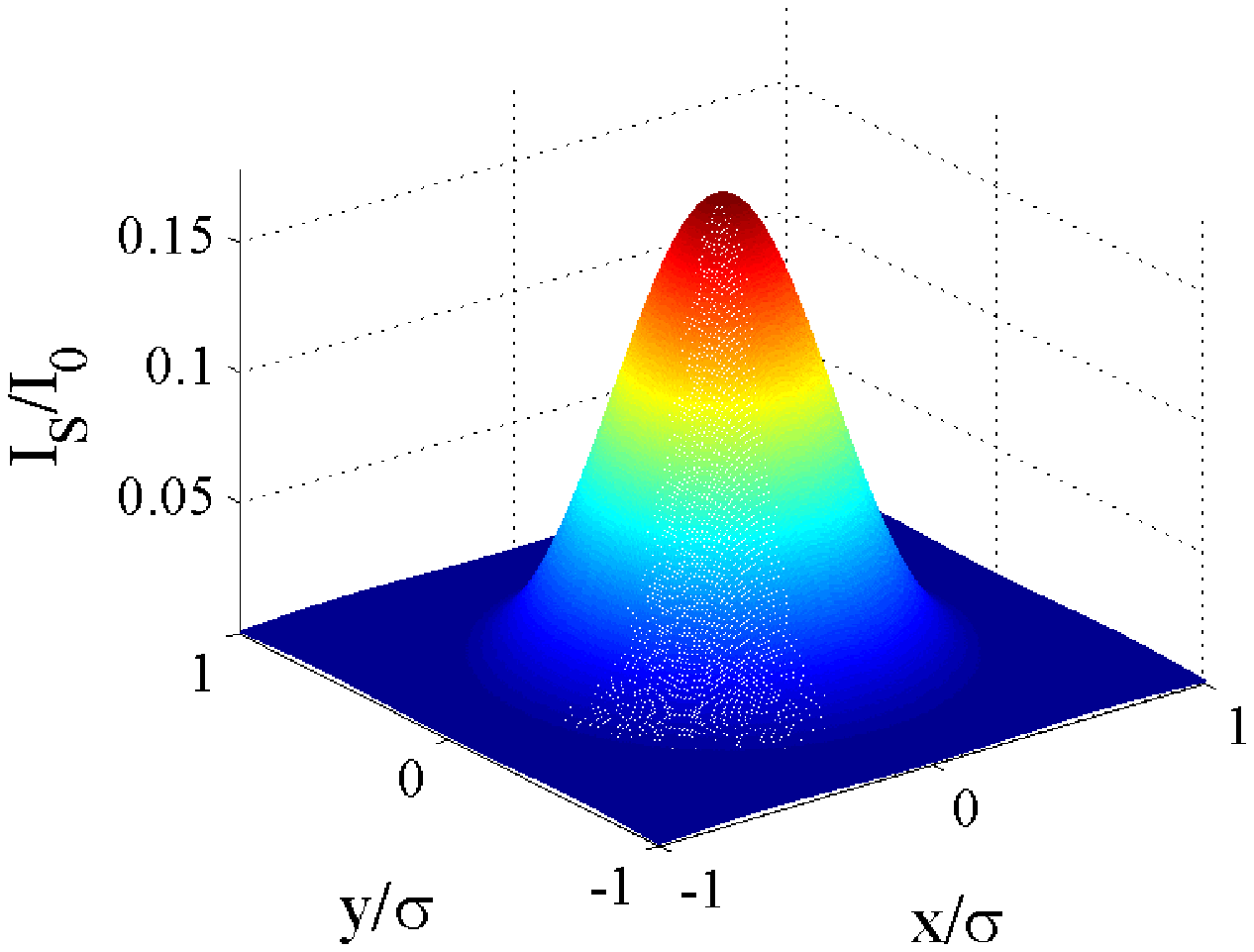}&\includegraphics[width=6cm]{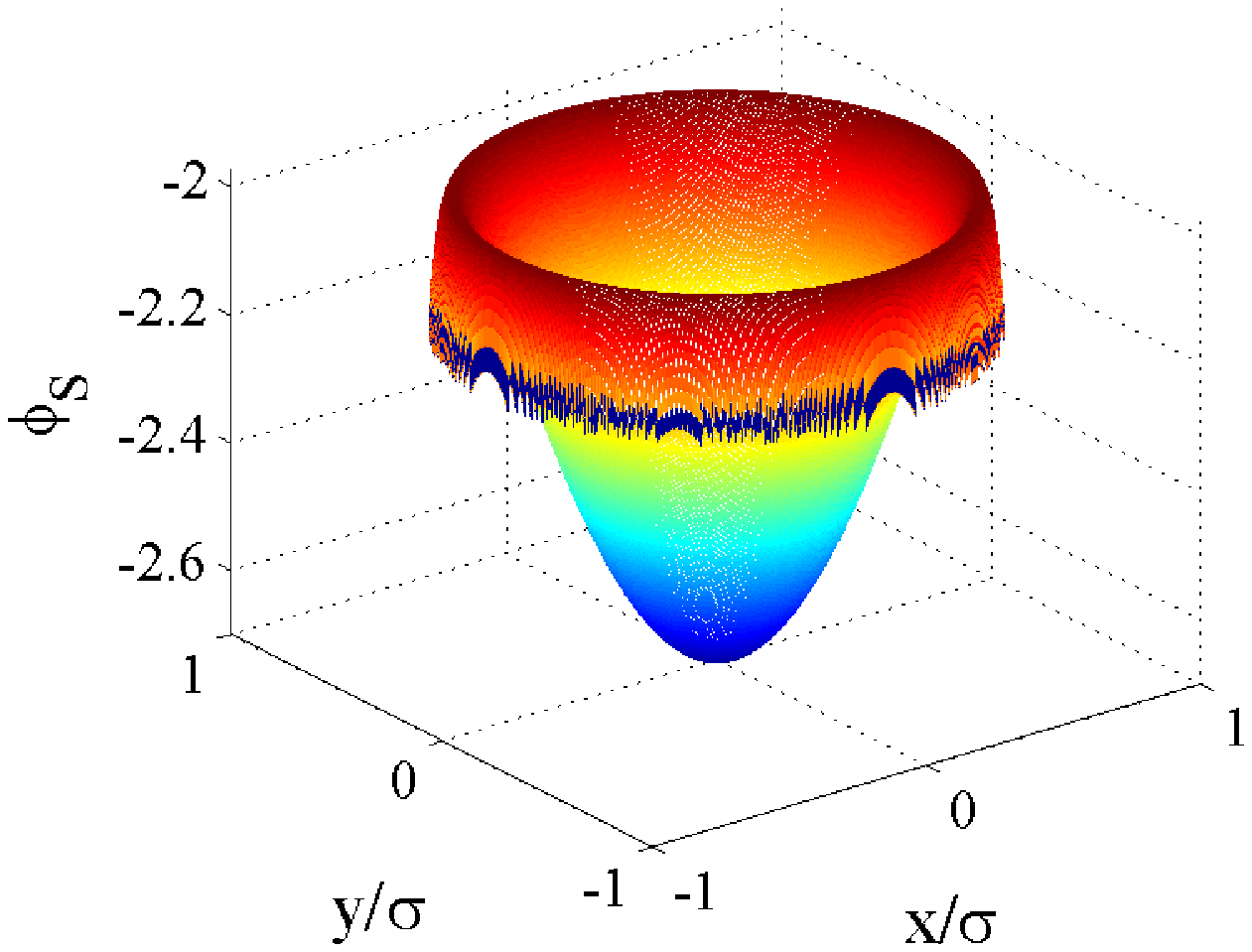}
\end{array}$
\caption{Transverse intensity profile (left) of the wave emitted
by the cloud, at $z=3\sigma$ and its wavefront (right).
Simulations realized using the analytic expression (\ref{Int2}),
for $N=10^4$, $\sigma=10$ and $\delta=10$.} \label{fig:WF}
\end{figure}

As for the emitted wave, it is concentrated in the forward
direction (see Fig.\ref{fig:WF}, left), and there is no
backscattering (not shown here). The wavefront phase does not
exhibit significant distortion in the central region of the
radiated beam (see Fig.\ref{fig:WF}, right).

\section{Conclusions}

In summary, we studied collective scattering from a dense and large atomic cloud with
gaussian density profile in terms of the eigenvalues of the interaction operator. This
enabled us to calculate the state generated by the interaction with a laser beam. We
found that this state considerably deviates from the 'timed Dicke' state at high optical densities.
In order to characterize this state, we calculated the phase front of the excited atomic dipole
moments and the force due to cooperative scattering by the atomic cloud.

Our approach consisted in expanding the solution on spherical harmonics, which form a complete
orthogonal basis of the Hermitian interaction matrix. Under the assumption that the system is
sufficiently dilute to neglect short-range effects due to dipole-dipole interactions, the continuous
spectrum approximation allows to transform infinite series into solvable integrals, and eventually to
derive analytical expressions for the most relevant observables, such as the scattered intensity and
radiation pressure force. These analytical expressions show good agreement with the numerical solution of
the $N$-body problem, and highlight the dependence on the accessible experimental parameters such as optical
thickness, atomic cloud size and laser frequency.

The analytical solution appears particularly useful for studying
the thermodynamic limit when $N\rightarrow\infty$ and
$V\rightarrow\infty$ with $N/V$ fixed, until collisions or non
resonant interactions come into play. The thermodynamic limit is
hardly accessible to $N$-body simulations, since the latter are
highly CPU consuming.

In contrast, the eigenvalue approach opens the possibility to
study the fascinating link between microscopic and macroscopic
domains of light scattering, and in particular between single-atom
scattering and Mie scattering for extended continuous samples
\cite{Prasad}.

For large optical thickness, the refraction index of the cloud
acts back on the driving field and shifts its phase. For this
reason, the solution for large optical thickness shows appreciable
deviations from that obtained assuming a symmetric timed Dicke
state for the atomic sample, since the latter corresponds to a
completely degenerate eigenvalue spectrum. The exact solution
(\ref{beta:ss}) takes into account the induced phase shift and
reduces to the symmetric timed Dicke state in the limit of
relatively small optical thickness.

An important further development of the present study should be to
understand how the dipole-dipole interactions contribute to the
observed cooperative effects, completing in this way the
cooperative scattering description for highly compressed and dense
atomic clouds.

\section{Acknowledgements}

We acknowledge fruitful discussions with R. Kaiser and T.
Bienaim\'{e}. We are also grateful to F. Staniscia for support for
the numerical simulations. The work has been supported by the
COSCALI network within the IRSES program of the European
Commission under grant number PIRSES-GA-2010-268717.

\appendix

\section{Derivation of the scattered electric field}

The scattered radiation is provided by the positive-frequency part
of the electric field operator
\begin{equation}\label{Es:a}
    \hat E_S(\mathbf{r},t)=\sum_{\mathbf{k}}{\cal E}_{k}\hat
    a_{\mathbf{k}}(t)
    e^{i\mathbf{k}\cdot \mathbf{r}-i\omega_k t}
\end{equation}
where ${\cal E}_k=(\hbar\omega_k/2\epsilon_0 V_{ph})^{1/2}$.
Integrating Eq.(\ref{aH}) with $a_{\mathbf{k}}(0)=0$, inserting it
in Eq.(\ref{Es:a}) and approximating the sum over the modes
$\mathbf{k}$ by an integral, we obtain
\begin{equation}\label{Es:1}
    \hat E_S(\mathbf{r},t)=-i\frac{V_{ph}}{8\pi^3}\sum_{m=1}^N
    \int_0^t dt'\hat\sigma_m(t-t')e^{i\omega_a t'}
    \int d\mathbf{k}\,{\cal E}_{k}g_k
    e^{i\mathbf{k}\cdot(\mathbf{r}-\mathbf{r}_m)-ickt'}.
\end{equation}
Introducing spherical coordinates, $d\mathbf{k}=dk k^2 d\phi
d\theta\sin\theta$, and integrating the angular part,
Eq.(\ref{Es:1}) becomes
\begin{equation}\label{Es:2}
    \hat E_S(\mathbf{r},t)=-\frac{V_{ph}}{4\pi^2}\sum_{m=1}^N\frac{1}{|\mathbf{r}-\mathbf{r}_m|}
    \int_0^t dt'\hat\sigma_m(t-t')e^{i\omega_a t'}
    \int_0^\infty dk k\,{\cal E}_{k}g_k\left\{
    e^{-ick(t'-|\mathbf{r}-\mathbf{r}_m|/c)}-e^{-ick(t'+|\mathbf{r}-\mathbf{r}_m|/c)}
    \right\}.
\end{equation}
Assuming the radiation spectrum centered around $k\approx k_0$, we
approximate $k{\cal E}_{k}g_k\approx k_0{\cal E}_{k_0}g_{k_0}$.
Then, extending the lower limit of integration of $k$ to
$-\infty$, we obtain for $t<|\mathbf{r}-\mathbf{r}_m|/c$
\cite{Prasad,Loudon}
\begin{equation}\label{Es:3}
    \hat E_S(\mathbf{r},t)\approx -\frac{dk_0^2}{4\pi\epsilon_0}\sum_{m=1}^N
    \frac{e^{ik_0 |\mathbf{r}-\mathbf{r}_m|}}{|\mathbf{r}-\mathbf{r}_m|}
    \hat\sigma_m(t-|\mathbf{r}-\mathbf{r}_m|/c).
\end{equation}
Neglecting the radiation retard in the limit $t\gg\sigma_R/c$,
where $\sigma_R$ is the cloud size, and approximating in the far
field limit $|\mathbf{r}-\mathbf{r}_m|\approx r-\hat{
\mathbf{n}}\cdot\mathbf{r}_m$ where $\hat{
\mathbf{n}}=\mathbf{r}/r$, Eq.(\ref{Es:3}) becomes
\begin{equation}\label{Es}
    \hat E_S(r,\theta,\phi,t)\approx
    -\frac{dk_0^2}{4\pi\epsilon_0
    r}e^{ik_0(r-ct)+i\Delta_0 t}\sum_{m=1}^N
    \hat\sigma_m(t) e^{-i\mathbf{k}_s\cdot\mathbf{r}_m},
\end{equation}
where
$\mathbf{k}_s=k_0(\sin\theta\cos\phi,\sin\theta\sin\phi,\cos\theta)$.
When applied on the state (\ref{state}), neglecting virtual
transitions, it yields $\hat E_S|\Psi\rangle={\cal E}_S|g_1\dots
g_N\rangle$ and where ${\cal E}_S$, in the continuous density
approximation will be given by Eq.(\ref{Es2}).

\section{Derivation of  Eq.(\ref{Force:fur:1})}

The angular integral in Eq.(\ref{Fez:cont}) is
\begin{equation}
\int_0^{2\pi}\mbox{d}\phi\int_0^\pi
  \mbox{d}\theta\sin\theta\cos\theta
  e^{-i\mathbf{k}\cdot(\mathbf{r}-\mathbf{r}')}=4\pi
  i\frac{z-z'}{|\mathbf{r}-\mathbf{r}'|}j_1(k_0|\mathbf{r}-\mathbf{r}'|).\label{id:j1}
\end{equation}
Since
\[
\frac{\partial}{\partial z}
j_0(k_0|\mathbf{r}-\mathbf{r}'|)=-k_0\frac{(z-z')}{|\mathbf{r}-\mathbf{r}'|}j_1(k_0|\mathbf{r}-\mathbf{r}'|),
\]
where $j_0(x)=\sin(x)/x$, Eq.(\ref{Fez:cont}) can be written as
\begin{equation}\label{Fe:2}
\langle\hat F_e\rangle=-i\frac{\hbar k_0\Gamma}{2 N}\int
\mbox{d}\mathbf{r} n(r)\left\{
\beta(\mathbf{r})\frac{\partial}{\partial (k_0 z)}\int
\mbox{d}\mathbf{r}'n(r')j_0(k_0|\mathbf{r}-\mathbf{r}'|)
\beta^*(\mathbf{r}') -\textrm{c.c.}\right\}.
\end{equation}
Using the expansions (\ref{sinc}) and (\ref{betaexp}) and
Eqs.(\ref{norm}) and (\ref{eigen}), we obtain
\begin{eqnarray}\label{exp2}
\int d\mathbf{r}' n(\mathbf{r}')
    j_0(k_0|\mathbf{r}-\mathbf{r}'|)\beta^*(\mathbf{r}')=\sum_{n=0}^\infty\sum_{m=-n}^{n}\alpha^{*}_{nm}\lambda_n
    j_n(k_0 r)Y^{*}_{nm}(\theta,\phi).
\end{eqnarray}
In spherical coordinates
\[
\frac{\partial}{\partial z}=\cos\theta\frac{\partial}{\partial
r}+\frac{\sin^2\theta}{r}\frac{\partial}{\partial \cos\theta}
\]
and
\begin{eqnarray}\label{Force:cont:2}
    \langle \hat F_{e}\rangle &=& -i\frac{\hbar k_0\Gamma}{2N}\int
\mbox{d}\mathbf{r} n(r)
    \left\{
    \beta(\mathbf{r})
    \sum_{n=0}^\infty\sum_{m=-n}^{n}\alpha^{*}_{nm}\lambda_n
    \left(
    \cos\theta\frac{\partial}{\partial
    (k_0 r)}+\frac{\sin^2\theta}{k_0 r}\frac{\partial}{\partial\cos\theta}
    \right)
    j_n(k_0 r)Y^{*}_{nm}(\theta,\phi)-\textrm{c.c.}\right\}
\end{eqnarray}
Still using Eq.(\ref{betaexp}),
\begin{eqnarray}\label{Force:cont:3}
    \langle\hat F_{e}\rangle
    &=& -\frac{\hbar k_0\Gamma}{2N}\sum_{p=0}^\infty\sum_{q=-p}^{p}
    \sum_{n=0}^\infty\sum_{m=-n}^{n}\lambda_n
    \int_0^\infty \mbox{d}r r^2 n(r)
    j_p(k_0 r)\int_0^{2\pi}\mbox{d}\phi\int_0^\pi \mbox{d}\theta\sin\theta\times\nonumber\\
    &\times &\left\{
    \frac{\partial j_n(k_0 r)}{\partial(k_0 r)}
    \cos\theta\,
    Y_{pq}(\theta,\phi)Y^{*}_{nm}(\theta,\phi)
    +\frac{j_n(k_0 r)}{k_0r}\sin^2\theta\,
    Y_{pq}(\theta,\phi)
    \frac{\partial Y^{*}_{nm}(\theta,\phi)}{\partial (\cos\theta)}
    \right\}i\left(\alpha_{pq}\alpha^{*}_{nm}-\textrm{c.c.}\right).
\end{eqnarray}
Assuming as before
\[
\alpha_{nm}=\alpha_n\delta_{m,0}
\]
and since
\[
Y_{n0}(\theta,\phi)=\sqrt{\frac{2n+1}{4\pi}}P_{n}(\cos\theta),
\]
Eq.(\ref{Force:cont:3}) becomes
\begin{eqnarray}\label{Force:cont:0}
    \langle\hat F_{e}\rangle
    &=& -\frac{\hbar k_0\Gamma}{4N}\sum_{p=0}^\infty
    \sum_{n=0}^\infty\lambda_n\sqrt{(2n+1)(2p+1)}
    \int_0^\infty \mbox{d}r r^2 n(r)
    j_p(k_0 r)\int_0^\pi d\theta\sin\theta\times\nonumber\\
    &\times &\left\{
    \frac{\partial j_n(k_0 r)}{\partial(k_0r)}
    \cos\theta\,
    P_{p}(\cos\theta)P_{n}(\cos\theta)
    +\frac{j_n(k_0 r)}{k_0r}\sin^2\theta\,
    P_{p}(\cos\theta)
    \frac{\partial P_{n}(\cos\theta)}{\partial (\cos\theta)}
    \right\}i\left(\alpha_{p}\alpha^{*}_{n}-\textrm{c.c.}\right).
\end{eqnarray}
Since
\[
\int_0^\pi d\theta\sin\theta\cos\theta\,
    P_{p}(\cos\theta)P_{n}(\cos\theta)=\int_{-1}^1 \mbox{d}x x P_{n}(x)P_{p}(x)
\]
and
\[
\int_0^\pi \mbox{d}\theta\sin^3\theta
    P_{p}(\cos\theta)
    \frac{\partial P_{n}(\cos\theta)}{\partial
    (\cos\theta)}=
    \int_{-1}^1 \mbox{d}x (1-x^2) P_{p}(x)\frac{\mbox{d}}{\mbox{d}x}P_{n}(x),
\]
using the identities
\[
(2p+1)xP_{p}(x)=(p+1)P_{p+1}(x)+pP_{p-1}(x),
\]
\[
(x^2-1)\frac{dP_{n}(x)}{dx}=n\left[x P_{n}(x)-P_{n-1}(x)\right]
\]
and
\[
\int_{-1}^1 \mbox{d}x P_{n}(x)P_{p}(x)=\frac{2}{2n+1}\delta_{n,p},
\]
we obtain
\[
\int_{-1}^1 \mbox{d}x x P_{n}(x)P_{p}(x)=\frac{2}{(2p+1)(2n+1)} \left\{
(p+1)\delta_{n,p+1}+p\,\delta_{p,n+1}\right\}
\]
and
\[
\int_{-1}^1 \mbox{d}x (x^2-1)
P_{p}(x)\frac{\mbox{d}}{\mbox{d}x}P_{n}(x)=\frac{2n}{(2n+1)(2p+1)} \left\{
p\,\delta_{p,n+1}-(n+1)\delta_{n,p+1} \right\}.
\]
By substituting these expressions in Eq.(\ref{Force:cont:0}) we
obtain
\begin{eqnarray}\label{Force:cont:1}
    \langle\hat F_{e}\rangle
    &=& -\frac{\hbar k_0\Gamma}{2N}\sum_{p=0}^\infty
    \sum_{n=0}^\infty i(\alpha_{p}\alpha^{*}_{n}-\textrm{c.c.})\frac{\lambda_n}{\sqrt{(2n+1)(2p+1)}}
    \int_0^\infty \mbox{d}r r^2 n(r)j_p(k_0 r)\times\nonumber\\
    &\times &\left\{
    (p+1)\left[\frac{\partial j_{p+1}(k_0 r)}{\partial(k_0r)}+(p+2)\frac{j_{p+1}(k_0
    r)}{k_0r}\right]\delta_{n,p+1}+
    p\left[\frac{\partial j_n(k_0 r)}{\partial(k_0r)}-n\frac{j_n(k_0
    r)}{k_0r}\right]\delta_{p,n+1}
    \right\}.
\end{eqnarray}
Since
\[
\frac{d j_{n+1}(z)}{d z}=j_n(z)-\frac{n+2}{z}j_{n+1}(z)
\]
and
\[
\frac{d j_{n}(z)}{d z}=-j_{n+1}(z)+è\frac{n}{z}j_{n}(z),
\]
using the definition (\ref{eigen}),
\begin{eqnarray}\label{Force:cont:4}
    \langle\hat F_{e}\rangle
    &=& -\frac{\hbar k_0\Gamma}{8\pi N}\sum_{p=0}^\infty
    \sum_{n=0}^\infty i(\alpha_{p}\alpha^{*}_{n}-\textrm{c.c.})\frac{\lambda_n\lambda_p}{\sqrt{(2n+1)(2p+1)}}
    \left\{
    (p+1)\delta_{n,p+1}-p\delta_{p,n+1}
    \right\}.
\end{eqnarray}
Eliminating one of the two sums, we obtain:
\begin{eqnarray}\label{Force:cont:5}
    \langle\hat F_{e}\rangle
    &=& -\frac{\hbar
    k_0\Gamma}{4\pi
    N}\sum_{n=0}^\infty\frac{(n+1)\lambda_n\lambda_{n+1}}{\sqrt{(2n+1)(2n+3)}}i(\alpha_{n}\alpha^{*}_{n+1}-\textrm{c.c.}).
\end{eqnarray}
In the stationary case,
\[
\alpha_n=\frac{\Omega_0}{\Gamma}\frac{i^n\sqrt{4\pi(2n+1)}}{2\delta+i(1+\lambda_n)}
\]
and the stationary force is
\begin{eqnarray}\label{Force:cont:ss:app}
    \langle F_{e}\rangle
    &=& -\hbar
    k_0\frac{2\Omega_0^2}{\Gamma N}\sum_{n=0}^\infty
    \frac{(n+1)\lambda_n\lambda_{n+1}[4\delta^2+(1+\lambda_n)(1+\lambda_{n+1})]}
    {[4\delta^2+(1+\lambda_n)^2][4\delta^2+(1+\lambda_{n+1})^2]}.
\end{eqnarray}

\end{document}